\documentclass[preprint,3p,12pt]{elsarticle}
\usepackage{tcolorbox}
\def\thedemobiblio#1{\smallskip\par
 \list{}{\labelwidth 0pt \leftmargin 1em \itemindent -1em \itemsep 1pt}
 \small \parindent 0pt
 \parskip 1.5pt plus .1pt\relax
 \def\newblock{\hskip .11em plus .33em minus .07em}
 \sloppy\clubpenalty4000\widowpenalty4000
 \sfcode`\.=1000\relax}

\newcommand{\bfg}[1]{\mbox{\boldmath $#1$}}

\newtheorem{theorem}{Theorem}[section]
\newenvironment{proof}[1][Proof]{\begin{trivlist}
		\item[\hskip \labelsep {\bfseries #1}]}{\end{trivlist}}

\usepackage{amssymb}
\usepackage{amsmath}
\usepackage{amsfonts}
\usepackage{amsbsy}
\usepackage{color}
\usepackage{float}
\usepackage{graphicx}
\usepackage{tcolorbox}
\usepackage{caption}
\usepackage{subcaption}
\usepackage{lineno}

\newtheorem{remark}{Remark}[section]

\journal{arXiv}
\begin{document}
\begin{sloppypar}

\title{\large {\bf
On Nonlocal Cohesive Continuum Mechanics and Cohesive Peridynamic Modeling (CPDM) of Inelastic Fracture
}
}

\author{Jing Han ${}^{\dag}$}
\author{Shaofan Li ${}^{\ddag}$\footnote{Email:shaofan@berkeley.edu}}
\author{Haicheng Yu ${}^{\S}$}
\author{Jun Li ${}^{\star}$}
\author{A-man Zhang ${}^{\dag}$}

\address{
${}^{\dag}$College Of Shipbuilding Engineering, Harbin Engineering University, Harbin, Heilongjiang, 150001, China,\\
${}^{\ddag}$ Department of Civil and Environmental Engineering,
University of California, Berkeley, \\
California, 94720, USA; \\
${}^{\S}$ College of Naval Architecture and Ocean Engineering, Dalian Maritime University, Dalian, Liaoning, 116026, China;\\
${}^{\star}$ College of Science, Wuhan University of Technology, Wuhan, Hubei, 430070, China.
}

\begin{abstract}
In this work, we developed a bond-based cohesive peridynamics model (CPDM)
and apply it to simulate inelastic
fracture by using the meso-scale Xu-Needleman cohesive potential \cite{Xu1993}.
By doing so, we have successfully developed a bond-based cohesive continuum mechanics
model with intrinsic stress/strain measures as well as consistent and built-in macro-scale
constitutive relations.
The main novelties of this work are:\\
(1) We have shown that the cohesive stress of the proposed nonlocal
cohesive continuum mechanics model is exactly the same as the
nonlocal peridynamic stress;\\
(2) For the first time, we have applied an irreversible built-in cohesive stress-strain
relation in a bond-based cohesive peridynamics to model inelastic material behaviors
without prescribing phenomenological
plasticity stress-strain relations;\\
(3) The cohesive bond force possesses both axial and tangential components,
and they contribute a nonlinear constitutive relation with variable
Poisson's ratios;\\
(4) The bond-based cohesive constitutive model is consistent with
the cohesive fracture criterion, and\\
(5) We have shown that the proposed method is able to model inelastic fracture
and simulate ductile fracture of small scale yielding in the nonlocal cohesive
continua.

Several numerical examples have been presented to be compared with the finite element based
continuum cohesive zone model, which shows that the proposed approach is a simple,
efficient and effective method to model inelastic fracture in the nonlocal
cohesive media.
\end{abstract}

\maketitle

\section{Introduction}
Peridynamics \cite{Silling2000,Silling2007,Silling2010,Silling2010L,madenci2014}
was originally proposed as a nonlocal reformulation of continuum mechanics aiming
at modeling fracture and damage in solids.
Peridynamics research has been an active research field in computational mechanics,
especially in numerical
simulation of fracture and failure in materials and structures
\cite{Warren2009,Breitenfeld2014Non}.
The non-local peridynamics theory is formulated with an integral form of
equation of motion,
which replaces the partial differential form of equation of motion in
conventional continuum mechanics of local form.
By doing so, it is applicable to a much broader class of displacement fields that
allow discontinuities and singularities, thus providing much needed physical modeling
of many non-local media such as cementitious concrete materials, soil and rocks,
ice and snow, and many other granular materials.

In spite of its success, the original bond-based peridynamics has some major limitations:
(1) It has been difficult to evaluate peridynamic stress in the bond-based peridynamics;
(2) Its main applications have been limited to model brittle fracture or crack growth in
macro-scale linear elastic solids with restrictions on certain material constants
such as Poisson's ratio;
(3) It needs a semi-empirical parameter, namely the critical bond stretch, $s_0$,
to set up the onset of fracture or crack growth criterion, and
(4) It has difficulty modeling material or structure fracture with
continuum mechanical stress and strain measures of finite deformation,
even though peridynamics is intrinsically formulated under the setting of
continuum mechanical finite deformation.

The main cause for these limitations is that the current formulation of the bond-based peridynamics
has not reached to a status to be a truly bond-based nonlocal continuum mechanics,
and these inadequacies are reflected
by  lacking of stress measures, corresponding macro-scale
constitutive models, as well as damage models or fracture criteria.
For example, Cauchy's relation is an intrinsic limitation for
the bond-based peridynamics for a fixed Poisson's ratio \cite{Trageser2020}.
Moreover, the peridynamic stress formulated by Lehoucq and Silling \cite{Lehoucq2008,Silling2010}
is cumbersome to use so that it has been rarely adopted in computations,
which leads to the lack of consistent macro-scale constitutive relations
in the bond-based peridynamics. In particular, the bond-based peridynamics almost does
not have an universally consistent inelastic constitutive
relation at macro-scale.

To address all these fundamental issues in the bond-based peridynamics,
in this work, we developed a bond-based cohesive peridynamic (CPDM) model for nonlocal
continua by utilizing
the meso-scale Xu-Needleman cohesive potential. In this paper, we shall demonstrate
that by combing the classical cohesive zone model \cite{Xu1994}.
The cohesive zone peridynamics has been studied by several authors, e.g.
\cite{Breitenfeld2014,Yang2020,Yang2021}, however, the focus of the present work is
not on cohesive zone peridynamics, but a general bond-based peridynamics
that utilizes the mesoscale cohesive potential to model
a nonlocal continuum. From this perspective, we
are developing a novel nonlocal continuum mechanics modeling.

The paper is organized into six sections. In Section 2, we first lay out
the kinematics of nonlocal continuum. Then, in Section 3, we present
the formal theory of cohesive nonlocal continuum.
One highlight of this work is the presentation of cohesive stress formulation,
which is elaborated in Section 4. Several numerical examples, both two-dimensional (2D)
and three-dimensional (3D), are presented in Section 5, to validate and verify
the proposed CPDM theory and formulation.
We summarize the work in Section 6 with a few remarks.

\section{Nonlocal continuum kinematics}

\begin{figure}[H]
	\begin{center}
		\includegraphics[height=3.3in]{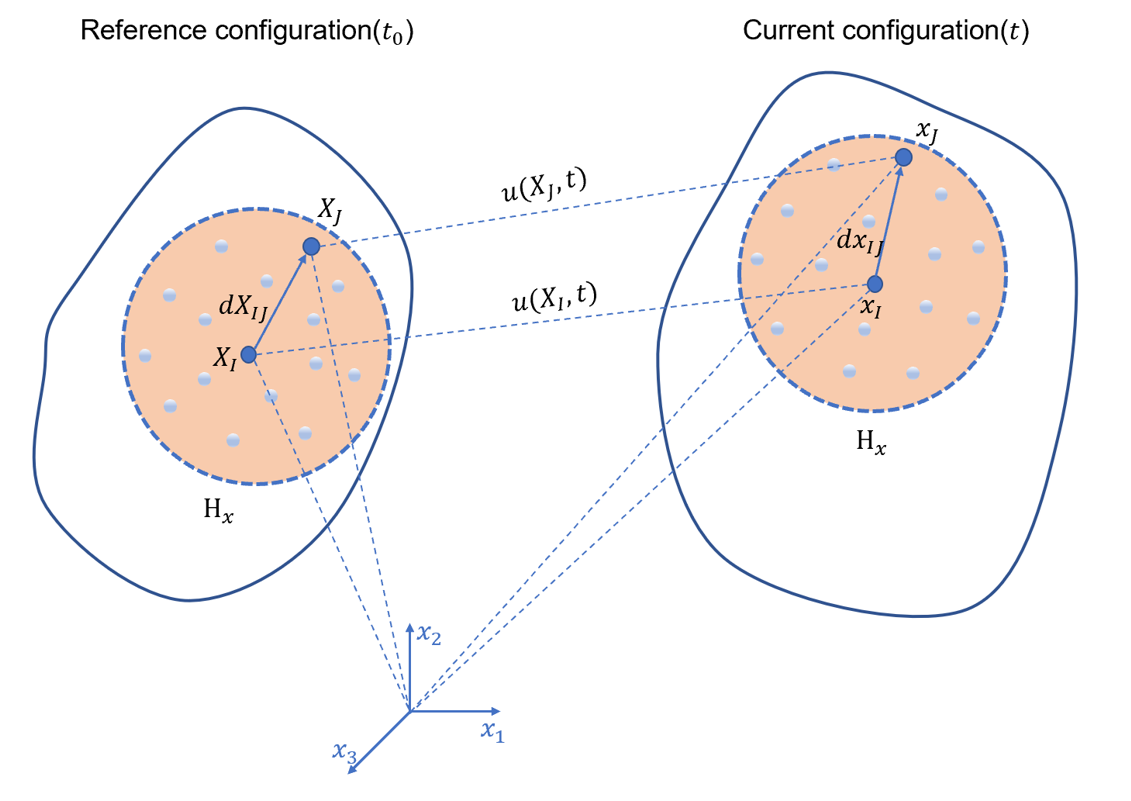}
	\end{center}
	\caption{Kinematics of material points}
	\label{fig:Illustration}
\end{figure}

To establish a bond-based peridynamics model for cohesive continua,
we first describe the material bond kinematics.
Given the referential and the current configurations $\mathcal{B}_0$ and $\mathcal{B}_t$,
for any pair of peridynamic particles $({\bf X}, {\bf X}^{\prime})$  that interact with each other,
the bond vector is described by $\boldsymbol{\xi}$ and $\boldsymbol{\eta}$ as follows,
\begin{equation}
\boldsymbol{\xi}=\boldsymbol{X}'-\boldsymbol{X}; ~~ \boldsymbol{\eta}=\boldsymbol{u}(\boldsymbol{X}',t)-\boldsymbol{u}(\boldsymbol{X},t),~{\rm and}~~
\boldsymbol{\zeta} = {\bf x}^{\prime} - {\bf x}~,
\label{eq:bond-describe}
\end{equation}
where ${\bf X}$ is the marker of the material point in the referential configuration $\mathcal{B}_0$,
while  ${\bf x}$ is the coordinate of the same material point in the current $\mathcal{B}_t$.
In Eq. (\ref{eq:bond-describe}), ${\bf u}({\bf X}, t)$ is the displacement
of the material point ${\bf X}$.The schematic diagram is demonstrated in Fig. \ref{fig:Illustration}.

Following the kinematic relation and notation of continuum mechanics, we have
\begin{equation}
{\bf x} = {\bf X} + {\bf u} ~\to~ \boldsymbol{\zeta} = \boldsymbol{\xi} + \boldsymbol{\eta}~.
\end{equation}
where $\boldsymbol{\xi}$ describes the original bond vector,
$\boldsymbol{\zeta}$ describes the deformed bond vector, or the bond vector in
the current configuration, while $\boldsymbol{\eta}$ is the deformation
of the bond vector.

We first define the first type nonlocal material gradient
for a vector function ${\bf G} ({\bf X})$  at
the point ${\bf X}$ as
\begin{equation}
\widetilde{\nabla}_X \otimes [ {\bf G} ({\bf X}) ]:= \Bigl[ \int_{\mathcal{H}_{X}} w(\xi)
\Delta {\bf G} \otimes ({\bf K}^{-1}{\bfg \xi}) d V_{{\bf X}^{\prime}} \Bigr]
\label{eq:NDoperator}
\end{equation}
where $\widetilde{\nabla}_X$ denotes
the nonlocal differential operator
in the referential configuration.
For more detailed discussions on nonlocal differential operators,
the readers are referred to \cite{Bergel2016,Kan2021}.
The integration domain in Eq.(\ref{eq:NDoperator}), $\mathcal{H}_{X}$,
is called as a horizon that is centered at ${\bf X}$, and
${\bfg \xi} ={\bf X}^{\prime} - {\bf X}$,
$\xi = |{\bfg \xi}|$ where
${\bf X}^{\prime}, {\bf X} \in \mathcal{H}_{X}$.
In the integrand of Eq.(\ref{eq:NDoperator}),
$\Delta {\bf G} :={\bf G}(\bf X ') - {\bf G}(\bf X)$,
and $w(\xi)$ is a window function or weight function,
which satisfies the condition
\[
\int_{\mathcal{H}_X} w(\xi) d V_{{\bf X}^{\prime}} = 1~.
\]
In practice, it is often chosen
as the Gaussian distribution function or the cubic spline
function.
The linear transformation
${\bf K}:= {\bf K}({\bf X})$ is the shape tensor or the moment
matrix at the material point ${\bf X}$ that is defined as
\begin{equation}
{\bf K} ({\bf X}):= \int_{\mathcal{H}_{X}} w(\xi) {\bfg \xi} \otimes
{\bfg \xi} d V_{{\bf X}^{\prime}}~.
\end{equation}

In actual computations, the nonlocal gradient of an arbitrary vector function ${\bf G} ({\bf X})$
may be calculated based on
the following formula,
\begin{equation}
\widetilde{\nabla}_X \otimes [ {\bf G} ({\bf X}_I) ]= \Bigl[ \sum_{J=1, J\not =I}^{N_I} w(X_{IJ})
\Delta {\bf G}_{IJ} \otimes ({\bf X}_{IJ} {\bf K}_I^{-1}) \Delta V_{J} \Bigr]
\end{equation}
where
\[
{\bf K}_I=
{\bf K} ({\bf X}_I)= \sum_{J=1, J\not =I}^{N_I} w(X_{IJ}) {\bf X}_{IJ} \otimes
{\bf X}_{IJ}  \Delta V_{J}~.
\]
and $V_J$ is the discrete volume associated with the particle $J$.

For example, we can write the nonlocal deformation gradient in a form as
\begin{tcolorbox}
\begin{eqnarray}
\widetilde{\mathbf{F}} ({\bf X}) &=& \widetilde{\nabla}_X {\bf x}
=\left[
\int_{\mathcal{H}_X} w(\xi){\bfg \zeta}
\otimes ({\bfg \xi} {\bf K}^{-1}_X) \mathrm{d} V_{{\bf X}^{\prime}} \right]~\to~
\nonumber
\\
\widetilde{\mathbf{F}} ({\bf X}_I) &=& \sum_{J=1}^N
w(\xi_{IJ}) {\bfg \zeta}_{IJ} \otimes {\bfg \xi}_{IJ} {\bf K}^{-1}({\bf X}_I) \Delta V_J
\end{eqnarray}
\end{tcolorbox}

Then in an abstract form, we may denote
the nonlocal gradient operator as a form of a local gradient operator
\begin{equation}
\widetilde{\nabla}_{{\bf X}} \otimes  (\bullet) \Bigm|_{{\bf X}} :=
\left[ \int_{\mathcal{H}_X} w(\xi)\Delta(\bullet)
\otimes ({\bfg \xi} {\bf K}_X^{-1}) {\rm d} V_{{\bf X}^{\prime}} \right]
\label{eq:DFF2}
\end{equation}
where the symbol $(\bullet) $ denotes the arbitrary vector field,
and $\Delta(\bullet):=(\bullet)^{\prime}-(\bullet) $. Note that Eq. (\ref{eq:DFF2})
defines the nonlocal differential operator by using linear function basis.
For higher order nonlocal differential operators theory,
readers may refer to \cite{Yan2020} and \cite{Yu2021} and references therein.

\section{Nonlocal cohesive continuum model}

Following Silling and Lequcq (2008), we have the nonlocal balance of linear
momentum as follows,
\begin{equation}
\rho ({\bf X}) \ddot{\bf u} ({\bf X}, t) = \int_{\mathcal{B}}
\Bigl(
{\bf t}^s ({\bf X}^{\prime}, {\bf X}, t) - {\bf t}^s ({\bf X}, {\bf X}^{\prime},t)
\Bigr) d V_{{\bf X}^{\prime}} + {\bf b}({\bf X},t)
\end{equation}
where $\rho$ is the mass density of the continuum medium;
${\bf b}({\bf X},t)$ is the body force per unit mass,
and
${\bf t}^s ({\bf X}^{\prime}, {\bf X}, t) $ is called the {\it force state vector},
and ${\bf f} ({\bf X}^{\prime}, {\bf X}) :=
{\bf t}^s ( {\bf X}^{\prime}, {\bf X}, t)- {\bf t}^s ({\bf X} , {\bf X}^{\prime}, t)$
represents the force density acting at
the material point ${\bf X}$ by the material point ${\bf X}^{\prime}$.

As shown by Silling and Lehoucq \cite{Silling2010},
the force density can be related to the force state vector as
\begin{equation}
{\bf t}^s ({\bf X}^{\prime}, {\bf X}, t) = {1 \over 2}
{\bf f} (\boldsymbol{\eta},\boldsymbol{\xi}),
~~{\rm and}~~
{\bf t}^s ({\bf X}, {\bf X}^{\prime}, t) = {1 \over 2}
{\bf f} (-\boldsymbol{\eta},-\boldsymbol{\xi}) ~.
\end{equation}
where ${\bf t}^s ({\bf X}^{\prime}, {\bf X})$
is the force state vector that material particle ${\bf X}^{\prime}$ exerts on the
material particle ${\bf X}$, in which the superscript indicates the force state;
$V_X$ is the volume of the particle ${\bf X}$ depending on the specific discretization,
while ${\bf f} (\boldsymbol{\eta},\boldsymbol{\xi})$ is a force density, which is required to
be antisymmetric, i.e.
\begin{equation}
{\bf f} (\boldsymbol{\eta},\boldsymbol{\xi}) = -
{\bf f} (-\boldsymbol{\eta},-\boldsymbol{\xi})~.
\end{equation}
In the literature, we often express the above property in an equivalent form,
\begin{equation}
{\bf f} ({\bf X}^{\prime}, {\bf X}) = - {\bf f} ({\bf X}, {\bf X}^{\prime})~.
\end{equation}

\subsection{Meso-scale Xu-Needleman model}

To construct the internal force density in a nonlocal cohesive continuum,
we adopt the mesoscale cohesive potential as the material bond potential, in
contrast with the prototype microelastic brittle (PMB) potential adopted in
the original bond based peridynamics, e.g. \cite{Silling2000}.

In this work, we adopt the mesoscale Xu-Needleman potential \cite{Xu1993}
as the material bond potential.
Unlike atomistic pair bond potential in molecular dynamics, the meso-scale Xu-Needleman
potential can generate both axial interaction force as well as tangential interaction
force.
To construct a pair bond with normal and tangential cohesive bond force components
we define
\begin{equation}
\boldsymbol{\eta}_n=(\boldsymbol{\eta}\cdot
\boldsymbol{n})\boldsymbol{n}~~{\rm and}~~ \boldsymbol{\eta}_t
=\boldsymbol{\eta}-(\boldsymbol{\eta}\cdot
\boldsymbol{n})\boldsymbol{n}
\label{eq:normal-tangential}
\end{equation}
where
\begin{equation}
\boldsymbol{n}=\frac{\boldsymbol{\xi}}{\left|\boldsymbol{\xi}\right|}
\label{eq:normal-direction}
\end{equation}
in other words
\begin{equation}
\boldsymbol{\eta}=\boldsymbol{\eta}_n+\boldsymbol{\eta}_t;
~~~ \boldsymbol{\eta}_n\perp\boldsymbol{\eta}_t;~~ \eta=\sqrt{\eta_n^2+\eta_t^2}~.
\label{eq:deltarelation}
\end{equation}

We consider the following meso-scale Xu-Needleman potential:
\begin{tcolorbox}
\begin{eqnarray}
\phi (\boldsymbol{\eta}) &=&
{\phi_n} \left \{
1 +
\displaystyle
\exp \bigl(
- {\frac{\bfg{\eta} \cdot {\bf n}}{\delta_n}
} \bigr)
\left\{
\bigl[
\displaystyle
1-r+ { \frac{\bfg{\eta} \cdot {\bf n}}{\delta_n}}
\bigr]
{\frac{1-q}{r-1}}
\right .
\right .
\nonumber
\\
&&
\left .
\left .
-\bigl[
\displaystyle
q+(\frac{r-q}{r-1})
{ \frac{\bfg{\eta} \cdot {\bf n}}{\delta_n} }
\bigr]
\exp \bigl(
-{\frac{1}{\delta_t^2} }
\Bigm|
{\bfg \eta} - ({\bfg \eta} \cdot {\bf n}) {\bf n}
\Bigm|^2
\bigr)
\right\} \right \}~.
\label{eq:needleman-potential-1}
\end{eqnarray}
\end{tcolorbox}
where $\phi_n$, $\delta_n$, $\delta_t$, $r$, and $q$ are coefficients
which will later be determined. The dimension of $\phi_n$ should be $N / m^5$,
$\delta_n$ and $\delta_t$ are characteristic lengths,
$r$ and $q$ are dimensionless. One can see that, the physical implications of
the five coefficients are not the same as those in original Xu-Needleman
potential. The bond force density that particle $\boldsymbol{X}'$ acts on
particle $\boldsymbol{X}$ can be obtained as follows:

\begin{equation}
\boldsymbol{f}=\frac{\partial\phi}{\partial\boldsymbol{\eta}}=\boldsymbol{f}_n+\boldsymbol{f}_t
\label{eq:phi-to-eta}
\end{equation}
Considering the fact
\begin{eqnarray}
&&{\frac{\partial}{\partial {\bfg \eta}}} \exp \bigl(
- {\frac{{\bfg \eta} \cdot {\bf n}}{\delta_n}}
\bigr) = - {1 \over \delta_n}
\exp \bigl(
- {{\bfg \eta} \cdot {\bf n} \over \delta_n}
\bigr) {\bf n}
\nonumber
\\
&&{\partial \over \partial {\bfg \eta}} \exp \Bigl(
-{1 \over \delta_t^2} \bigm|
{\bfg \eta}- {\bfg \eta} \cdot {\bf n}
\bigm|^2 \Bigr)
= - {2 \over \delta_t^2}
\exp \Bigl(
-{1 \over \delta_t^2} \bigm|
{\bfg \eta}- {\bfg \eta} \cdot {\bf n}
\bigm|^2 \Bigr)
\Bigl(
{\bfg \eta} - ({\bfg \eta} \cdot {\bf n}){\bf n}
\Bigr)
\nonumber
\end{eqnarray}
we then have
\begin{eqnarray}
\boldsymbol{f}(\boldsymbol{X},\boldsymbol{X}') &=&
-\frac{\phi_n}{\delta_n}\exp{(-\frac{\boldsymbol{\eta}\cdot\boldsymbol{n}}{\delta_n})}\{
[(-r+\frac{\boldsymbol{\eta}\cdot\boldsymbol{n}}{\delta_n})\frac{1-q}{r-1}\nonumber\\
&-&[q+(\frac{r-q}{r-1})\frac{\boldsymbol{\eta}\cdot\boldsymbol{n}}{\delta_n}]\exp(-\frac{1}{\delta_t^2}\left|\boldsymbol{\eta}-(\boldsymbol{\eta}\cdot\boldsymbol{n})\boldsymbol{n}\right|^2)\nonumber\\
&+&(\frac{r-q}{r-1})\exp(-\frac{1}{\delta_t^2}\left|\boldsymbol{\eta}-(\boldsymbol{\eta}\cdot\boldsymbol{n})\boldsymbol{n}\right|^2)]\boldsymbol{n}\nonumber\\
&-&[\frac{2\delta_n}{\delta_t^2}(q+(\frac{r-q}{r-1})\frac{\boldsymbol{\eta}\cdot\boldsymbol{n}}{\delta_n})\nonumber\\
&\cdot&\exp(-\frac{1}{\delta_t^2}\left|\boldsymbol{\eta}-(\boldsymbol{\eta}\cdot\boldsymbol{n})\boldsymbol{n}\right|^2)](\boldsymbol{\eta}-(\boldsymbol{\eta}\cdot\boldsymbol{n})\boldsymbol{n})\}
\label{eq:phi-to-eta-1}
\end{eqnarray}
It should be reminded that, when deriving $\boldsymbol{f}(\boldsymbol{X},\boldsymbol{X}')$, the local
coordinate system established on origin $\boldsymbol{X}$ is adopted; in contrast, the derivation of $\boldsymbol{f}(\boldsymbol{X}',\boldsymbol{X})$ is based on the coordinate system whose origin is $\boldsymbol{X}'$. Then the bond force that particle $\boldsymbol{X}$ acts on particle $\boldsymbol{X}'$
is as follows:
\begin{eqnarray}
\boldsymbol{f}(\boldsymbol{X}',\boldsymbol{X}) &=&
-\frac{\phi_n}{\delta_n}\exp{(-\frac{\boldsymbol{\eta}\cdot\boldsymbol{n}}{\delta_n})}\{
[(-r+\frac{\boldsymbol{\eta}\cdot\boldsymbol{n}}{\delta_n})\frac{1-q}{r-1}\nonumber\\
&-&[q+(\frac{r-q}{r-1})\frac{\boldsymbol{\eta}\cdot\boldsymbol{n}}{\delta_n}]\exp(-\frac{1}{\delta_t^2}\left|\boldsymbol{\eta}-(\boldsymbol{\eta}\cdot\boldsymbol{n})\boldsymbol{n}\right|^2)\nonumber\\
&+&(\frac{r-q}{r-1})\exp(-\frac{1}{\delta_t^2}\left|\boldsymbol{\eta}-(\boldsymbol{\eta}\cdot\boldsymbol{n})\boldsymbol{n}\right|^2)](-\boldsymbol{n})\nonumber\\
&-&[\frac{2\delta_n}{\delta_t^2}(q+(\frac{r-q}{r-1})\frac{\boldsymbol{\eta}\cdot\boldsymbol{n}}{\delta_n})\nonumber\\
&\cdot&\exp(-\frac{1}{\delta_t^2}\left|\boldsymbol{\eta}-(\boldsymbol{\eta}\cdot\boldsymbol{n})\boldsymbol{n}\right|^2)](-\boldsymbol{\eta}+(\boldsymbol{\eta}\cdot\boldsymbol{n})\boldsymbol{n})\}
\label{eq:phi-to-eta-2}
\end{eqnarray}

Equation (\ref{eq:phi-to-eta-2}) proves that the bond interaction $\boldsymbol{f}$ is anti-symmetric. Considering the unloading process, we present the following scalar values of the normal and tangential components of bond interaction:

\begin{equation}
f_n = \left \{
\begin{array}{lcl}
\displaystyle
{\phi_n \over \delta_n} \exp \Bigl(
- {{\bfg \eta} \cdot {\bf n} \over \delta_n}
\Bigr)
\left \{
{{\bfg \eta} \cdot {\bf n} \over \delta_n} \exp \Bigl(
- {1 \over \delta_t^2} |{\bfg \eta} - ({\bfg \eta} \cdot {\bf n}){\bf n}|^2 \Bigr)
\right .
&&
\\
\\
\displaystyle
 \left .
+ {1 - q \over r-1}
\left [
1 - \exp \Bigl(
- {1 \over \delta_t^2} |{\bfg \eta} - ({\bfg \eta} \cdot {\bf n}){\bf n}|^2 \Bigr)
\right ]
\left (
r - { {\bfg \eta} \cdot {\bf n} \over \delta_n}
\right )
\right \},&{\rm if}&~\eta_n \ < \ \eta_{n, max}~{\rm and}~\dot{\eta}_n > 0
\\
\\
\displaystyle
{f_{n,max} \over \eta_{n,max}} \eta_n, &{\rm if}&
\eta_n  < \eta_{n,max}~{\rm and}~\dot{\eta}_n <0
\end{array}
\right .
\end{equation}
where $f_{n,max}= f_n (\eta_{n,max}), \dot{\eta}_n >0$; and
\begin{equation}
f_t = \left \{
\begin{array}{l}
\displaystyle
{2 \phi_n } \exp \Bigl(
- {\bfg{\eta} \cdot {\bf n} \over \delta_n}
\Bigr)
{\left| {\bfg \eta} - (\bfg{\eta} \cdot {\bf n}) {\bf n} \right| \over \delta_t^2}
\Bigl(
q + \Bigl( {r -q \over r-1} \Bigr){\bfg{\eta} \cdot {\bf n} \over \delta_n}
\Bigr)
\\
\\
\cdot
\exp \Bigl(
- {1 \over \delta_t^2} |{\bfg \eta} - ({\bfg \eta} \cdot {\bf n}){\bf n}|^2 \Bigr),
~~{\rm if}~~\eta_t < \eta_{t,max}~{\rm and}~\dot{\eta}_t > 0
\\
\\
\displaystyle
{f_{t,max} \over \eta_{t,max}} \eta_t, ~~~~~~~~~~~~~~~~{\rm if}~~ \eta_t <
\eta_{t,max}~{\rm and}~\dot{\eta}_t <0
\end{array}
\right .
\end{equation}
where $f_{t,max}= f_t (\eta_{t,max}), \dot{\eta}_t >0$.

We define the normal direction as the direction along the bond between particles $I$ and $J$,
while the tangential direction is defined as the direction perpendicular to the normal direction.
In two-dimensional cases,
there is only one tangential direction.
In three-dimensional cases,
there is a plane perpendicular to the normal direction,
 in which we can define two mutually perpendicular tangential directions as
 ${\bf t}_{s1}$ and ${\bf t}_{s2}$ respectively.
In computations, we still consider
one tangential direction $ {\bf t}_{t}$,
and this direction is determined as
\[
{\bf t} = { {\bfg \eta} - ({\bfg \eta} \cdot {\bf n}) {\bf n}
\over
| {\bfg \eta} - ({\bfg \eta} \cdot {\bf n}) {\bf n}|
 }~.
\]

In other words, ${\bf t}_{t}$ is
the direction of the resultant force of ${\bf t}_{s1}$ and ${\bf t}_{s2}$.
The cases for 2D and 3D are shown in Fig. 2. Figure \ref{fig:deltan} shows the Xu-Needleman
cohesive laws in the normal and tangential directions
as a function of $\eta_{n}$ and $\eta_{t}$.
\begin{figure}
	\centering
	\begin{subfigure}[t]{0.4\linewidth}
	\centering
	\includegraphics[height=1.9in]{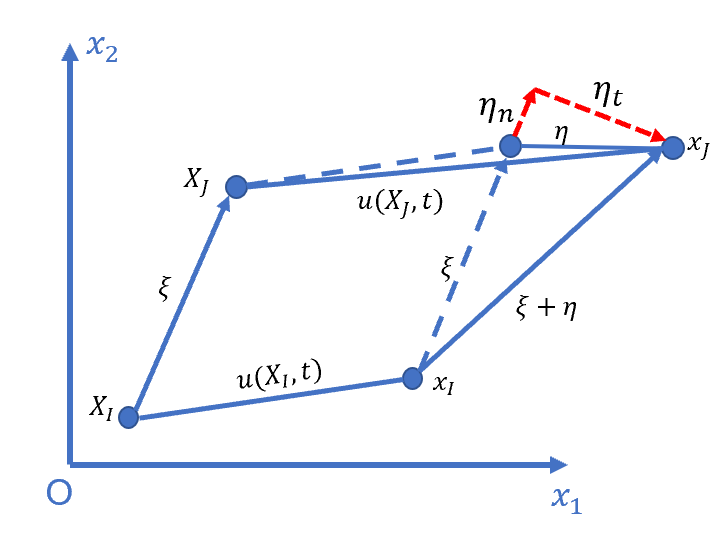}
	\begin{center}
     (a)
    \end{center}	
    \end{subfigure}
    \begin{subfigure}[t]{0.4\linewidth}
    	\centering
    	\includegraphics[height=2.0in]{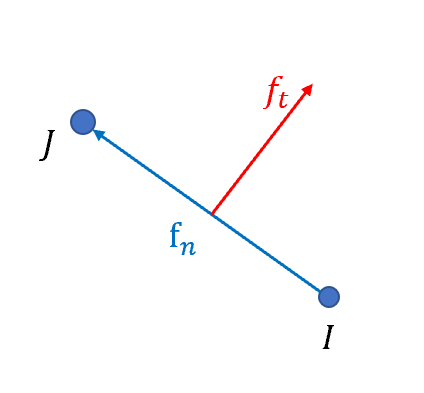}
    \begin{center}
     (b)
    \end{center}	
    \end{subfigure}
    \\
	\centering
\begin{subfigure}[t]{0.4\linewidth}
	\centering
	\includegraphics[height=2in]{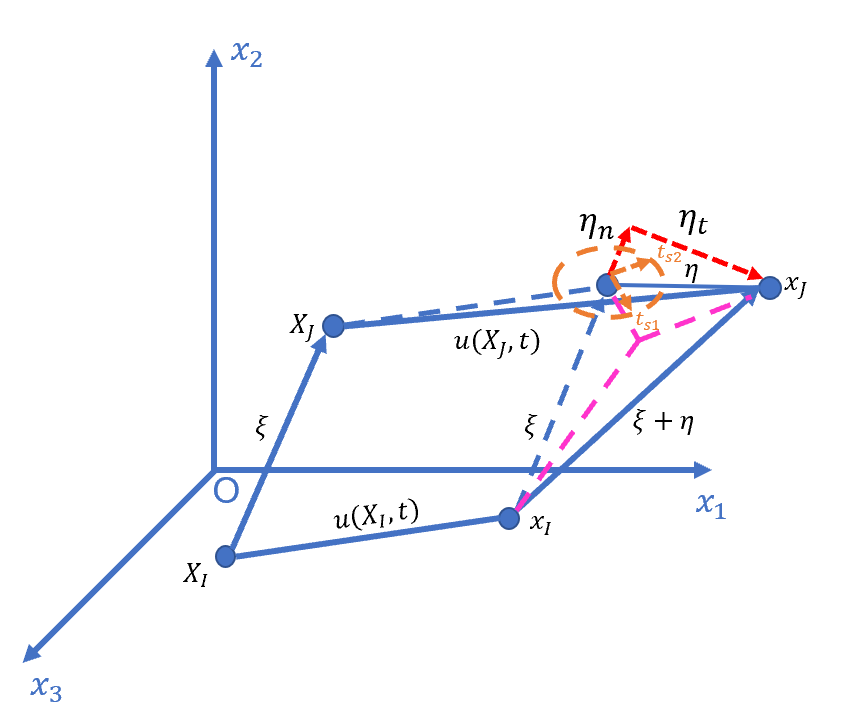}
	\begin{center}
		(c)
	\end{center}	
\end{subfigure}
\begin{subfigure}[t]{0.4\linewidth}
	\centering
	\includegraphics[height=2.0in]{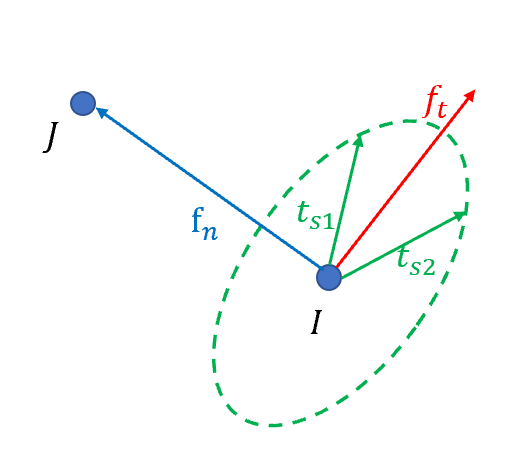}
	\begin{center}
		(d)
	\end{center}	
\end{subfigure}
    \caption{Tangential and normal stretches of a mesoscale pair bond:
    (a)(b) Two-dimensional case, and (c)(d)  Three-dimensional case.}
    \label{fig:direcctions}
\end{figure}

Adopting the Cauchy-Born rule, we assume that
in a horizon centered at $\boldsymbol{X}$, the following relation holds:
\begin{equation}
\boldsymbol{\zeta }=\boldsymbol{F}\cdot\boldsymbol{\xi}
\label{eq:general-1}
\end{equation}
where $\boldsymbol{F}$ is the deformation gradient at ${\bf X}$,
which is a constant two-point tensor
in the entire horizon.
Thus, Eq. (\ref{eq:general-1}) leads to the following equations:
\begin{eqnarray}
 \boldsymbol{F} \cdot \boldsymbol{\xi} &=&
\boldsymbol{\xi} + \boldsymbol{\eta}~~\to~~
\frac{\partial \boldsymbol{\eta}}{\partial \boldsymbol{F}}
 = \boldsymbol{I}^{(2)} \otimes \boldsymbol{\xi}
\label{eq:general-2}
\end{eqnarray}
in which $\boldsymbol{I}^{(2)}$ is the second order unit tensor.
\begin{figure}
	\centering
	\begin{subfigure}[t]{0.4\linewidth}
	\centering
	\includegraphics[width=2.41in]{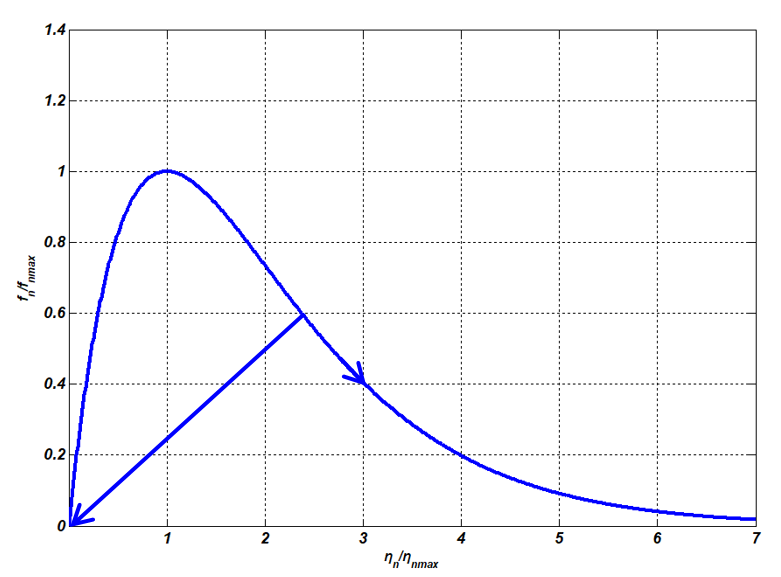}
	\begin{center}
		(a)
	\end{center}
   \end{subfigure}
	\begin{subfigure}[t]{0.4\linewidth}
	\centering
	\includegraphics[width=2.45in]{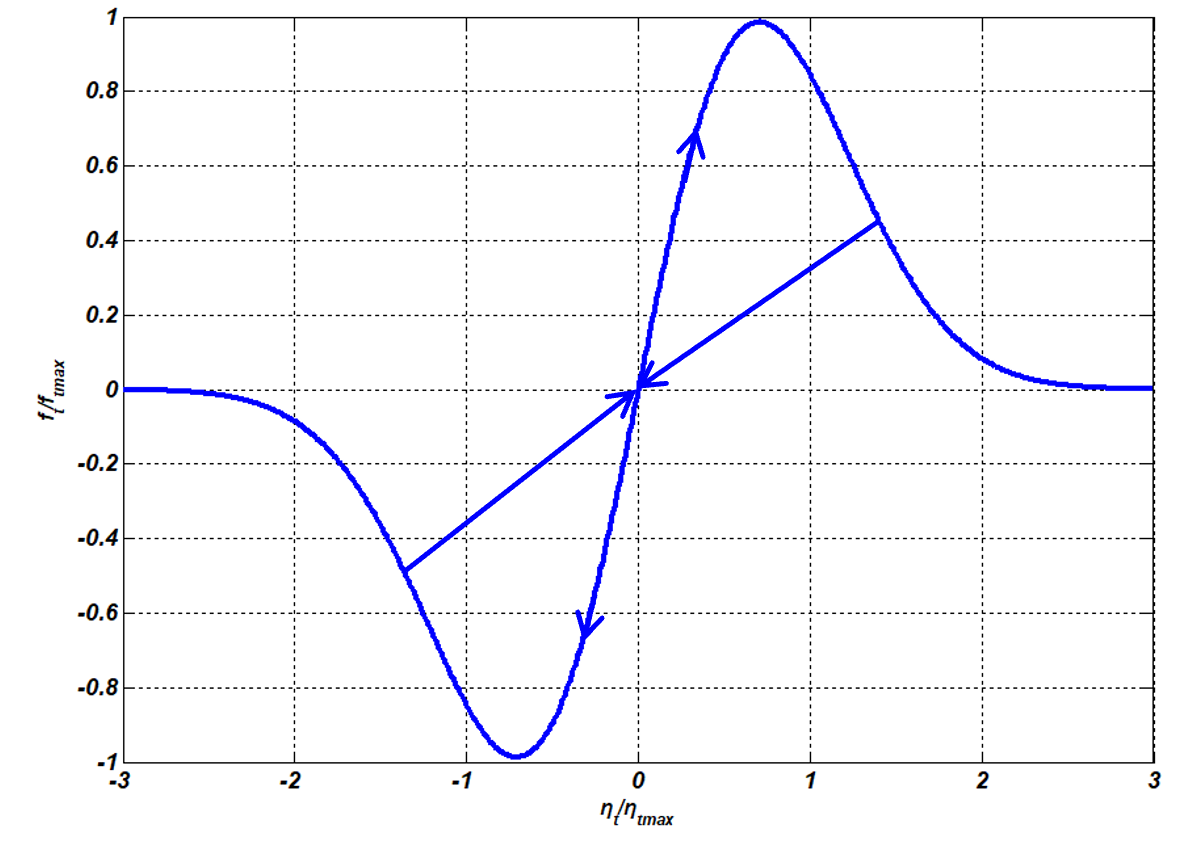}
	\begin{center}
		(b)
	\end{center}
\end{subfigure}
	\caption{The Xu-Needleman cohesive model:
(a) Normal cohesive law $f_n (\eta_n) $ while $\eta_{t}=0$ , and
(b) Shear cohesive law $f_t (\eta_t) $ while $\eta_{n}=0$.}
	\label{fig:deltan}
\end{figure}

If only considering the elastic range of the cohesive medium,
we can find the second derivative of $\phi$ as
\begin{eqnarray}
{\partial^2 \phi \over \partial {\bfg \eta} \partial {\bfg \eta}}
&=& - {1 \over \delta_n} {\bf n} \otimes {\partial \phi \over \partial {\bfg \eta}}
-(\frac{1-q}{r-1}){\phi_n \over \delta_n^2}
\exp \bigl( - {{\bfg \eta} \cdot {\bf n} \over {\delta_n}} \bigr) \boldsymbol{n} \otimes \boldsymbol{n}
\nonumber\\
&+&\frac{2q\phi_n}{\delta_t^2} \exp{(-\frac{\boldsymbol{\eta}\cdot\boldsymbol{n}}{\delta_n}
-\frac{1}{\delta_t^2}\left|\boldsymbol{\eta}
-(\boldsymbol{\eta}\cdot\boldsymbol{n})\boldsymbol{n})\right|^2)}[-\boldsymbol{\eta} \otimes\nonumber\\
&&(\frac{1}{\delta_n}\boldsymbol{n}+\frac{2}{\delta_t^2}(\boldsymbol{\eta}
-(\boldsymbol{\eta}\cdot\boldsymbol{n})\boldsymbol{n}))+\boldsymbol{I}^{(2)}]\nonumber\\
&-&\frac{2q\phi_n}{\delta_t^2}\exp{(-\frac{\boldsymbol{\eta}\cdot\boldsymbol{n}}{\delta_n}
-\frac{1}{\delta_t^2}\left|\boldsymbol{\eta}-(\boldsymbol{\eta}\cdot\boldsymbol{n})
\boldsymbol{n})\right|^2)}\boldsymbol{n}\otimes[ \nonumber\\
&&\boldsymbol{n}-\frac{\boldsymbol{\eta}\cdot\boldsymbol{n}}{\delta_n}\boldsymbol{n}
-\frac{2\boldsymbol{\eta}\cdot\boldsymbol{n}}{\delta_t^2}(\boldsymbol{\eta}
-(\boldsymbol{\eta}\cdot\boldsymbol{n})\boldsymbol{n})] \nonumber\\
&+&\frac{\phi_n}{\delta_n}(\frac{r-q}{r-1})\exp{(-\frac{\boldsymbol{\eta}
\cdot\boldsymbol{n}}{\delta_n}-\frac{1}{\delta_t^2}\left|\boldsymbol{\eta}
-(\boldsymbol{\eta}\cdot\boldsymbol{n})\boldsymbol{n})\right|^2)}\boldsymbol{n}\otimes[ \nonumber\\
&&\frac{1}{\delta_n}\boldsymbol{n}+\frac{2}{\delta_t^2}(\boldsymbol{\eta}
-(\boldsymbol{\eta}\cdot\boldsymbol{n})\boldsymbol{n})]+\frac{2\phi_n}{\delta_n\delta_t^2}
(\frac{r-q}{r-1})\cdot \nonumber\\
&&\exp{(-\frac{\boldsymbol{\eta}\cdot\boldsymbol{n}}{\delta_n}
-\frac{1}{\delta_t^2}\left|\boldsymbol{\eta}
-(\boldsymbol{\eta}\cdot\boldsymbol{n})\boldsymbol{n})\right|^2)}
\{\boldsymbol{\eta}\otimes[\boldsymbol{n}-\frac{\boldsymbol{\eta}
\cdot\boldsymbol{n}}{\delta_n}\boldsymbol{n}- \nonumber\\
&&\frac{2\boldsymbol{\eta}\cdot\boldsymbol{n}}{\delta_t^2}(\boldsymbol{\eta}
-(\boldsymbol{\eta}\cdot\boldsymbol{n})\boldsymbol{n})]+(\boldsymbol{\eta}
\cdot\boldsymbol{n})\boldsymbol{I}^{(2)}\}-\frac{2\phi_n}{\delta_n\delta_t^2}(\frac{r-q}{r-1})\cdot \nonumber\\
&&\exp{(-\frac{\boldsymbol{\eta}\cdot\boldsymbol{n}}{\delta_n}
-\frac{1}{\delta_t^2}\left|\boldsymbol{\eta}
-(\boldsymbol{\eta}\cdot\boldsymbol{n})\boldsymbol{n})\right|^2)}\boldsymbol{n}\otimes[ \nonumber\\
&&2(\boldsymbol{\eta}\cdot\boldsymbol{n})\boldsymbol{n}
-\frac{(\boldsymbol{\eta}\cdot\boldsymbol{n})^2}{\delta_n}\boldsymbol{n}
-\frac{2(\boldsymbol{\eta}\cdot\boldsymbol{n})^2}{\delta_t^2}(\boldsymbol{\eta}
-(\boldsymbol{\eta}\cdot\boldsymbol{n})\boldsymbol{n})]
\end{eqnarray}

Now we can define the strain energy density as follows
\begin{tcolorbox}
\begin{equation}
W(\boldsymbol{X})=\frac{1}{2V_\mathcal{H}}\int_{\mathcal{H}}\int_{\mathcal{H}}\phi(\boldsymbol{\eta},\boldsymbol{\xi})
d \boldsymbol{X}'' d \boldsymbol{X}'
\label{eq:general-5}
\end{equation}
where $\mathcal{H}$ is the horizon of $\boldsymbol{X}$, $V_{\mathcal{H}}$ is the volume of horizon.
\end{tcolorbox}

We can then derive the first Piola-Kirchhoff stress tensor
at $\boldsymbol{X}$ as
\begin{equation}
\boldsymbol{P}(\boldsymbol{X})=\frac{\partial
W(\boldsymbol{X})}{\partial \boldsymbol{F}}=\frac{1}{2V}
\int_{\mathcal{H}} \int_{\mathcal{H}}
\frac{\partial \phi(\boldsymbol{\eta},\boldsymbol{\xi})}
{\partial \boldsymbol{\eta}}\cdot
\frac{\partial \boldsymbol{\eta}}{\partial \boldsymbol{F}}
d\boldsymbol{X}''d\boldsymbol{X}'
\label{eq:general-6}
\end{equation}
where
\begin{equation}
\frac{\partial \boldsymbol{\eta}}{\partial \boldsymbol{F}}
 = \boldsymbol{I}^{(2)} \otimes \boldsymbol{\xi}~.
\label{eq:eta-to-F}
\end{equation}
\medskip

Substituting Eq. (\ref{eq:phi-to-eta})
and Eq. (\ref{eq:eta-to-F}) into Eq. (\ref{eq:general-6}),
we obtain the expression of the cohesive stress as follows,
\\
\begin{tcolorbox}
\begin{equation}
\boldsymbol{P}(\boldsymbol{X}) =
\frac{\partial W(\boldsymbol{X})}{\partial \boldsymbol{F}}
=
\frac{1}{2V_{\mathcal{H}}}\int_{\mathcal{H}} \int_{\mathcal{H}}
(\boldsymbol{f} \otimes \boldsymbol{\xi})
d \boldsymbol{X}''d \boldsymbol{X}'~
\label{eq:GEN9}
\end{equation}
\end{tcolorbox}

We can rewrite Eq. (\ref{eq:GEN9}) as
\begin{equation}
\boldsymbol{P}(\boldsymbol{X}) =
\frac{1}{2}\int_{\mathcal{H}}
\bar{\boldsymbol{f}} \otimes \boldsymbol{\xi} d \boldsymbol{X}'~
\label{eq:general-10}
\end{equation}
where
\begin{equation}
\bar{\bf f} := {1 \over V_{\mathcal{H}}}
\int_{\mathcal{H}} {\bf f} ({\bf X}, {\bf X}^{\prime}) d {\bf X}^{\prime}
\end{equation}

To find the macroscale elasticity tensors corresponding
to the mesoscale Xu-Needleman potential, we can compute
\begin{equation}
\boldsymbol{C}(\boldsymbol{X})=
\frac{\partial \boldsymbol{P}(\boldsymbol{X})}{\partial\boldsymbol{F}}
=\frac{1}{2}
\int_{\mathcal{H}}\frac{\partial}{\partial\boldsymbol{F}}
(\frac{\partial\phi}{\partial\boldsymbol{F}})d\boldsymbol{\xi}
=\frac{1}{2}\int_{\mathcal{H}}
\frac{\partial}{\partial\boldsymbol{F}}(\bar{\boldsymbol{f}}
\otimes\boldsymbol{\xi})dV_X
\label{eq:elasticitytensor-mod-1}
\end{equation}

To evaluate Eq. (\ref{eq:elasticitytensor-mod-1}), one needs to carry out
double integrations. For simplicity, we may assume that
the force density is continuous and smooth in the interior
of the material domain, so that when the size of the horizon is small enough
we can adopt the following approximation,
\begin{equation}
\bar{\bf f} ({\bf X}) := {1 \over V_{\mathcal{H}}}
\int_{\mathcal{H}} {\bf f} ({\bf X}, {\bf X}^{\prime}) d {\bf X}^{\prime}
\approx {1 \over V_X}
\int_{\mathcal{V}_X} {\bf f} ({\bf X}, {\bf X}^{\prime}) d {\bf X}^{\prime}
\label{eq:fbar}
\end{equation}
where $V_X$ is an infinitesimal volume that contains the material point ${\bf X}$
i.e. the center of the horizon $\mathcal{H}$. By continuity of ${\bf f}({\bf X})$, we then have
\begin{equation}
\bar{\bf f} ({\bf X})
\approx {1 \over V_X}
\int_{\mathcal{V}_X} {\bf f} ({\bf X}, {\bf X}^{\prime}) d {\bf X}^{\prime} = {\bf f} ({\bf X}),~~~
{\rm as}~~V_X \to 0~.
\end{equation}

By replacing the nonlocal force density to the local force density, we can
obtain the explicit expression of the elasticity tensor $\boldsymbol{C}(\boldsymbol{X})$,
we first instead consider another fourth order tensor $\boldsymbol{C}'(\boldsymbol{X})$ as follows:
\begin{eqnarray}
\boldsymbol{C}'(\boldsymbol{X})&=&
\frac{1}{2}\int_{\mathcal{H}}(\frac{\partial\boldsymbol{f}}{\partial\boldsymbol{F}}
\otimes\boldsymbol{\xi})d\boldsymbol{\xi}
= \frac{1}{2}\int_{\mathcal{H}}
(\frac{\partial\boldsymbol{f}}{\partial\boldsymbol{\eta}}\cdot
{\partial {\bfg \eta} \over \partial {\bf F}}
\otimes\boldsymbol{\xi})dV_X
\nonumber
\\
&=&
\frac{1}{2}\int_{\mathcal{H}} \Bigl(
\frac{\partial\boldsymbol{f}}{ \partial {\bfg \eta}}
\otimes {\bfg \xi}
\otimes\boldsymbol{\xi} \Bigr)dV_X
\label{eq:elasticitytensor-1}
\end{eqnarray}

The relationship between $\boldsymbol{C}(\boldsymbol{X})$ and $\boldsymbol{C}'(\boldsymbol{X})$ is as follows:
\begin{equation}
{C}_{ijkl}(\boldsymbol{X})={C}'_{mnst}(\boldsymbol{X})\delta_{mi}\delta_{nk}\delta_{sl}\delta_{tj}
\label{eq:elasticitytensor-mod-2}
\end{equation}

When ${\bfg \eta} =\boldsymbol{0}$, $\phi(\boldsymbol{0}) =0$ and considering
\begin{equation}
{\bf f} ={\partial \phi \over \partial {\bfg \eta}} (\boldsymbol{0}) = \boldsymbol{0}
\end{equation}
finally we have
\begin{tcolorbox}
\begin{eqnarray}
{\bf C}' (\boldsymbol{0}) &=& {1 \over 2 } \int_{\mathcal{H}}
\Bigl(
{\phi_n \over \delta_n^2}- {2 \phi_n q \over \delta_t^2}
\Bigr )
{ {\bfg \xi} \otimes {\bfg \xi} \otimes {\bfg \xi} \otimes {\bfg \xi}
\over |{\bfg \xi}|^2} d V_X
\nonumber
\\
&&+
{1 \over 2 } \int_{\mathcal{H}} {2 \phi_n q \over \delta_t^2}
{\bf I} \otimes {\bfg \xi} \otimes {\bfg \xi} d V_X~.
\end{eqnarray}
\end{tcolorbox}

\begin{remark}
By replacing the nonlocal force density with the local force density
is an analog of the Cauchy-Born in crystalline solids.
Without such approximation, one may still be able to find the macroscale
elasticity tensor of the nonlocal medium. However, its value may be
different, because of taking into account of the nonlocal interaction effect.
\end{remark}

\smallskip
\subsection{Macroscale material constants for the Xu-Needleman potential}
Considering spherical horizon and denoting the radius of the horizon as $H$, we then have
\[
\Omega_X = {4 \pi \over 3} H^3~.
\]
We can the explicitly evaluate the following integral
\begin{eqnarray}
&&\int_{\mathcal{H}} { {\bfg \xi} \otimes {\bfg \xi} \otimes {\bfg \xi} \otimes {\bfg \xi}
\over |{\bfg \xi}|^2} d V
= \bigl(\int_0^H r^4 d r \bigr) \int_{\mathcal{H}} {\bf n} \otimes {\bf n} \otimes {\bf n} \otimes {\bf n}
d \omega
\nonumber
\\
&& \to~
 \bigl(\int_0^H r^4 d r \bigr)
\bigl(
\int_{S_2} n_m n_n n_s n_t d \omega
\bigr)=\bigl( {H^5 \over 5} \bigr) \Bigl( {4 \pi \over 15} \Bigr)
\Bigl( \delta_{mn} \delta_{st} + \delta_{ms} \delta_{nt} + \delta_{mt} \delta_{ns}
\Bigr)
\end{eqnarray}
and
\begin{eqnarray}
&&\int_{\mathcal{H}} {\bf I} \otimes {\bfg \xi} \otimes {\bfg \xi} d V
=  \Bigl(\int_0^H r^4 d r \Bigr)
\int_{S_2} {\bf I} \otimes {\bf n} \otimes {\bf n} d \omega
\nonumber
\\
&& \to~
\Bigl(\int_0^H r^4 d r \Bigr)
\int_{S_2} \delta_{mn} n_s n_t d \omega
=\bigl( { H^5 \over 5} \bigr)
\Bigl( {4 \pi \over 3} \Bigr)  \delta_{mn} \delta_{st}
\end{eqnarray}

Thus for three-dimensional nonlocal solids, we have
\begin{equation}
C'_{mnst} = \frac{4\pi H^3}{3}\left\{{H^2 \over 50}
\Bigl(
{\phi_n \over \delta_n^2} - {2 \phi_n q \over \delta_t^2}
\Bigr)
\Bigl(
\delta_{mn} \delta_{st} + \delta_{ms} \delta_{nt} + \delta_{mt} \delta_{ns}
\Bigr)
+
{H^2 \over 10}
\Bigl(
{2 \phi_n q \over \delta_t^2}
\Bigr) \delta_{mn} \delta_{st}
\right\}
\end{equation}
and
\begin{equation}
C_{ijkl} = \frac{4\pi H^3}{3}\left\{{H^2 \over 50}
\Bigl(
{\phi_n \over \delta_n^2} - {2 \phi_n q \over \delta_t^2}
\Bigr)
\Bigl(
\delta_{ik} \delta_{lj} + \delta_{il} \delta_{kj} + \delta_{ij} \delta_{kl}
\Bigr)
+
{H^2 \over 10}
\Bigl(
{2 \phi_n q \over \delta_t^2}
\Bigr) \delta_{ik} \delta_{lj}\right\}
\end{equation}

In particular, we can then find that
\begin{equation}
C_{1111} = \frac{4\pi H^3}{3}{H^2 \phi_n \over 50} \Bigl( { 3 \over \delta_n^2}
+ { 4 q \over \delta_t^2}
\Bigr),~~C_{1122} = \frac{4\pi H^3}{3}{H^2\phi_n \over 50}
\Bigl(
{1 \over \delta_n^2} - {2 q \over \delta_t^2}
\Bigr)
\end{equation}
where $q=\phi_t/\phi_n$.

For isotropic materials, we have
\begin{eqnarray}
 C_{1111} &=& {E \over (1+\nu) (1-2\nu) } (1-\nu),
\\
 C_{1122} &=& {E \over (1+\nu) (1-2\nu) } \nu,
\end{eqnarray}

Then $\phi_n$ and $q$ can be accordingly solved as follows:
\begin{eqnarray}
 q &=& {{1-4\nu} \over {2(1+\nu)}}{{\delta_t^2} \over {\delta_n^2}},
\\
 \phi_n &=& {10E\delta_n^2 \over {H^2(1-2\nu)\frac{4\pi H^3}{3}}},
\end{eqnarray}
where the Poisson's ratio must obey the constraint $\nu<1/4$.

In two-dimensional cases,
the plane strain problems should have the same formulations as that of the three-dimensional case.
Now, we consider the case of plane stress problems, in which
the horizon has the volume
 \[
 \Omega_X= \pi H^2 B;
 \]
 where $B$ is the thickness of the planar plate.
 Thus, one can derive that
 \begin{eqnarray}
 &&\int_{\mathcal{H}} { {\bfg \xi} \otimes {\bfg \xi} \otimes {\bfg \xi} \otimes {\bfg \xi}
\over |{\bfg \xi}|^2} d V
= B\bigl(\int_0^H r^3 d r \bigr) \int_{\mathcal{H}} {\bf n} \otimes {\bf n} \otimes {\bf n} \otimes {\bf n}
d \omega
\nonumber
\\
&& \to~
  B \Bigl(\int_0^H r^3 d r \Bigr)
\int_{S_1} n_m n_n n_s n_t  d \theta =
\Bigl( { BH^4 \over 4} \Bigr)
{\pi \over 4} \bigl(
\delta_{mn} \delta_{st} + \delta_{ms} \delta_{nt}
+ \delta_{mt} \delta_{ns}
\bigr),
 \end{eqnarray}
 and
 \begin{eqnarray}
 &&\int_{\mathcal{H}} {\bf I} \otimes {\bfg \xi} \otimes {\bfg \xi} d V
=  B\Bigl(\int_0^H r^3 d r \Bigr)
\int_{S_2} {\bf I} \otimes {\bf n} \otimes {\bf n} d \omega
\nonumber
\\
&& \to~
B \Bigl(\int_0^H r^3 d r \Bigr)
\int_{S_1} \delta_{mn} n_s n_t  d \theta
=\Bigl( { BH^4 \over 4} \Bigr)
\pi  \delta_{mn} \delta_{st}~.
\end{eqnarray}

These lead to
\begin{equation}
C'_{mnst} = \pi H^2B\left\{{H^2 \phi_n \over 32}
\Bigl(
{1 \over \delta_n^2} - {2  q \over \delta_t^2}
\Bigr)
\Bigl(
\delta_{mn} \delta_{st} + \delta_{ms} \delta_{nt} + \delta_{mt} \delta_{ns}
\Bigr)
+
{H^2 \over 8}
\Bigl(
{2 \phi_n q \over \delta_t^2}
\Bigr) \delta_{mn} \delta_{st}\right\}
\end{equation}
and
\begin{equation}
C_{ijkl} = \pi H^2B\left\{{H^2 \phi_n \over 32}
\Bigl(
{1 \over \delta_n^2} - {2  q \over \delta_t^2}
\Bigr)
\Bigl(
\delta_{ik} \delta_{lj} + \delta_{il} \delta_{kj} + \delta_{ij} \delta_{kl}
\Bigr)
+
{H^2 \over 8}
\Bigl(
{2 \phi_n q \over \delta_t^2}
\Bigr) \delta_{ik} \delta_{lj}\right\}
\end{equation}

In particular, we have
\[
C_{1111} =
\pi H^2B{H^2 \phi_n \over 32}
\Bigl(
{ 3\over \delta_n^2} + {2  q \over \delta_t^2}
\Bigr)~~~{\rm and}~~
C_{1122} =
\pi H^2B{H^2 \phi_n \over 32}
\Bigl(
{ 1 \over \delta_n^2} - {2 q \over \delta_t^2}
\Bigr)~.
\]

For isotropic materials under the plane stress condition, we have
\[
C_{1111}={E \over (1-\nu^2)},
\]
and
\[
C_{1122}={E\nu \over (1-\nu^2)},
\]
based on which $q$ and $\phi_n$ can be accordingly obtained:
\begin{eqnarray}
q &=& {{1-3\nu} \over {2(1+\nu)}}{\delta_t^2 \over {\delta_n^2}},
\\
\phi_n &=& {8E\delta_n^2 \over {H^2(1-\nu)\pi H^2B}},
\end{eqnarray}
where the Poisson's ratio must satisfy the condition $\nu<1/3$.
By comparing with the original bond-based peridynamics formulation,
for the cohesive peridynamics model the nonlocal Poisson's ratio is variable,
even though it is subjected an upper-bound constraint.

\begin{remark}
The above relations suggest that the mesoscale pair-wise Xu-Needleman potential defies
the Cauchy relation --- a setback suffered for almost all pair-wise atomistic potentials.
This is because that the Xu-Needleman potential
offers both tension bond and shear bond simultaneously, making it
a suitable candidate in nonlocal cohesive continuum modeling.
\end{remark}

\subsection{
A Smith-Ferrante type cohesive model
}

For nonlocal cohesive media under finite deformation,
we can also introduce
the following Smith-Ferrante type potential function \cite{Ortiz1999}
as an alternative mesoscale potential for the nonlocal cohesive continuum,
which  provides a universal binding potential that can be also written as,
\begin{equation}
\phi(\boldsymbol{\eta},\boldsymbol{\xi})=\phi_n\sigma_c
e\eta_c
\left[1-\bigl(1+
\frac{\left|\boldsymbol{\eta}\right|}{\eta_c} \bigr)
\exp \bigl( -\frac{\left|\boldsymbol{\eta}\right|}{\eta_c} \bigr)
\right]~.
\label{eq:micropotential-2}
\end{equation}
and its corresponding force equals to
\begin{equation}
{\bf f} (\boldsymbol{\eta})=\frac{\partial\phi}{\partial {\bfg \eta}}
=\phi_n \sigma_c
\exp \Bigl(1 -\frac{\left|\boldsymbol{\eta}\right|}{\eta_c}\Bigr)
\frac{\boldsymbol{\eta}}{\eta_c}
\end{equation}
where $\eta_c = |{\bfg \eta}_c|$ is the critical value,
and when ${\bfg \eta} = {\bfg \eta}_c$ the bond force,
\[
{\bf f} = {\partial \phi \over \partial {\bfg \eta}}
\Bigm|_{{\bfg \eta}={\bfg \eta}_c} ~\to~{\bf t}_{max}
\]
reaches its peak value.

In the elastic range, we
can also derive the expression of the bond force vector as follows,
\begin{equation}
{\bf f} (\boldsymbol{\eta})=\frac{\partial\phi}{\partial {\bfg \eta}}
=\phi_n \sigma_c
\exp \Bigl(1 -\frac{\left|\boldsymbol{\eta}\right|}{\eta_c}\Bigr)
\frac{\boldsymbol{\eta}}{\eta_c}
\end{equation}

The magnitude of the bond force $f=|{\bf f}|$
can be defined as
\begin{equation}
f=\frac{\partial\phi}{\partial\eta}=
 \phi_n e\sigma_c\frac{\eta}{\eta_c}\exp \bigl({-\frac{\eta}{\eta_c}}\bigr)
\label{eq:bondforce-1}
\end{equation}
and the normal and tangential components of the bond force can be derived in the following:

\begin{equation}
f_n=\frac{\partial\phi}{\partial\eta_n}=
\frac{\phi_n \eta_n \sigma_c \exp \bigl(1-\frac{\eta}{\eta_c}\bigr)}
{\eta_c};
~{\rm and}~~ ~f_t=\frac{\partial\phi}{\partial\eta_t}=\frac{\phi_n \eta_t
\sigma_c \exp \bigl(1-\frac{\eta}{\eta_c}\bigr)}{\eta_c}~.
\label{eq:bondforcecomponent}
\end{equation}

It is straightforward to show that:
\begin{equation}
\boldsymbol{f}=\boldsymbol{f}_n+\boldsymbol{f}_t
\end{equation}

The derivative of the scalar amplitude bond force relative to $\eta$ is:
\begin{equation}
\frac{df}{d\eta}=-{\phi_n \over \eta_c^2}
\bigl(
e\sigma_c
\exp \bigl( {-\frac{\eta}{\eta_c}}(\eta-\eta_c)\bigr)~.
\label{eq:dbondforce-1}
\end{equation}

Equation (\ref{eq:dbondforce-1}) indicates that
the bond force $f$ reaches its maximum when $\eta=\eta_c$.
We expediently assume that for the case where $\eta\le\eta_c$,
the material is in elastic phase, and when $\eta\ge\eta_c$ the material
is in inelastic phase. Also, this conclusion is invariant under coordinate transformation.
\begin{figure}[H]
	\begin{center}
		\includegraphics[height=2.3in]{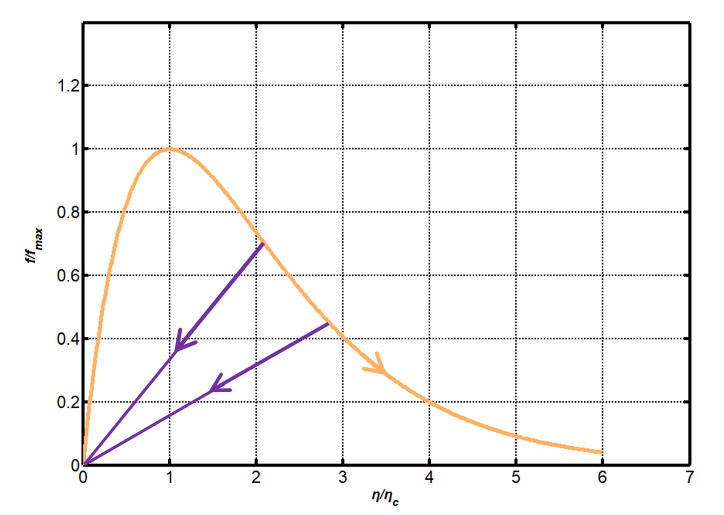}
	\end{center}
	\caption{Loading-unloading relation for the Smith-Ferrante cohesive media}
		\label{fig:loading}
\end{figure}

Considering the nonlocal strain energy density as follows
\begin{equation}
W(\boldsymbol{X})=\frac{1}{2 V_{\mathcal{H}}}
\int_{\mathcal{H}} \int_{\mathcal{H}}
\phi(\boldsymbol{\eta},\boldsymbol{\xi})
d {\bf X}^{\prime} d {\bf X}^{\prime \prime}
\end{equation}
we can then derive the first Piola-Kirchhoff stress tensor
at the location of ${\bf X}$ as
\begin{eqnarray}
\boldsymbol{P}(\boldsymbol{X})&=&
\frac{\partial W(\boldsymbol{X})}{\partial \boldsymbol{F}}
=\frac{1}{2V_{\mathcal{H}}} \int_{\mathcal{H}}\int_{\mathcal{H}}
\frac{\partial \phi(\boldsymbol{\eta},\boldsymbol{\xi})}{\partial |\boldsymbol{\eta}|}
\frac{\partial \boldsymbol{|\eta}|}{\partial \boldsymbol{\eta}}
\frac{\partial \boldsymbol{\eta}}{\partial \boldsymbol{F}}
d {\bf X}^{\prime} d {\bf X}^{\prime \prime}
\nonumber
\\
&=&
\frac{1}{2 V_{\mathcal{H}}}
\int_{\mathcal{H}}
\int_{\mathcal{H}}
\Bigl[
\phi_n \sigma_c
\exp \Bigl( 1 -\frac{|\boldsymbol{\eta}|}{|\boldsymbol{\eta}|_c}
\Bigr)
{\boldsymbol{\eta} \otimes \boldsymbol{\xi}
\over |\boldsymbol{\eta}|_c}
\Bigr]
d {\bf X}^{\prime} d {\bf X}^{\prime \prime}
\nonumber
\\
&=&
\frac{1}{2}\int_{\mathcal{H}}
\Bigl[
\bar{\bf f} \otimes \boldsymbol{\xi}
\Bigr]
d {\bf X}^{\prime}~
\label{eq:general-101}
\end{eqnarray}

Accordingly, we can find the elasticity tensor
for Smith-Ferrante type cohesive continuum media as\\
\begin{tcolorbox}
\begin{eqnarray}
\mathbb{C}_{SF} (\boldsymbol{X}) &=&
{\partial^2 W \over \partial {\bf F} \partial {\bf F}}
= {\partial \boldsymbol{P}({\bf X}) \over \partial \boldsymbol{F}}
\nonumber
\\
&\approx&
\frac{1}{2}
\int_{\mathcal{H}}
\frac{\phi_n\sigma_c e^{1-\frac{|\boldsymbol{\eta}|}
{|\boldsymbol{\eta}|_c}}}{ |\boldsymbol{\eta}|_c}
\Bigl(
\boldsymbol{I}^{(2)}
\otimes {\boldsymbol{\xi}} \otimes \boldsymbol{\xi}
-
{
\boldsymbol{\eta} \otimes \boldsymbol{\eta}
\over
 |\boldsymbol{\eta}| |\boldsymbol{\eta}|_c
 }
 \otimes  {\boldsymbol{\xi}} \otimes \boldsymbol{\xi}
 \Bigr)
d V_{\boldsymbol{X}}
\label{eq:general-14}
\end{eqnarray}
\end{tcolorbox}

One may find that the initial elastic tensor is given as
\begin{equation}
\mathbb{C}_{SF} ({\bf 0})=
\frac{1}{2}
\int_{\mathcal{H}}
\frac{\phi_n \sigma_c}{\eta_c}
\Bigl(
\boldsymbol{I}^{(2)}
\otimes {\boldsymbol{\xi}} \otimes \boldsymbol{\xi}
 \Bigr)
d V_{\boldsymbol{X}},
\label{eq:general-15}
\end{equation}
which does not possess the initial shear modulus.
This is because the Smith-Ferrante potential is
an atomistic pair bond potential that does not have
tangential bond displacement initially.

\subsection{Determination of characteristic lengths}
One of distinguished features of the cohesive continuum is
its internal length scale.
For the Xu-Needleman potential,
$\delta_{n}$ and $\delta_{t}$ are two characteristic length scales
that are defined as the maximum elastic bond stretches, i.e.
\begin{eqnarray}
\delta_{n}(\boldsymbol{\xi}) &=& \left|\boldsymbol{\xi}\right|c_{n},
\\
\delta_{t}(\boldsymbol{\xi})&=&\left|\boldsymbol{\xi}\right|c_{t},
\label{eq:characteristic-length-1}
\end{eqnarray}
where $c_n$ and $c_t$ are the two maximum elastic strains for normal and tangential
deformations of a pair bond, respectively.
Comparing with the treatment in \cite{Xu1993},
we can also define two critical bond strains or stretches
$s_n$ and $s_t$ for determining the critical or maximum stretches of the bond: $\delta_{nc}$ and $\delta_{tc}$:
\begin{eqnarray}
\delta_{nc}(\boldsymbol{\xi}) &=& \left|\boldsymbol{\xi}\right|s_{n},
\\
\delta_{tc}(\boldsymbol{\xi}) &=& \left|\boldsymbol{\xi}\right|s_{t},
\label{eq:characteristic-length-1-1}
\end{eqnarray}
where $s_n$ and $s_t$ are the critical bond strains or stretches
before the bond is broken.

Before, we determine the critical stretches, we first note that
the cohesive elastic potential of a Xu-Needleman bond may be interpreted as
the elastic bond energy when the bond force reaches to the peak loading forces,
which can be obtained as follows,
\begin{eqnarray}
\phi_{Ie} &=&
{\phi_n} \left \{
1 +
\displaystyle
\exp \bigl(
- {1
} \bigr)
\left\{
\bigl[
\displaystyle
1-r+ {1}
\bigr]
{\frac{1-q}{r-1}}
-\bigl[
\displaystyle
q+(\frac{r-q}{r-1})
\bigr]
\right\} \right \}=
\frac{e-2}{e}\phi_n
~
\label{eq:characteristic-length-2}
\end{eqnarray}
and
\begin{eqnarray}
\phi_{IIe} &=&
{\phi_n} \left \{
1 +
\displaystyle
\left\{
\bigl[
\displaystyle
1-r
\bigr]
{\frac{1-q}{r-1}}
\displaystyle
-q
\exp \bigl(
-1
\bigr)
\right\} \right \}=\frac{e-1}{e}q\phi_n~
\label{eq:characteristic-length-3}
\end{eqnarray}

To determine the critical stretches, we adopt the criteria of
the critical energy release.
It is assumed that the critical energy releases are achieved when all the bonds
connecting to the center particle of a given horizon reach their
corresponding critical stretches.
We assume that
the critical stretch of the normal deformation of pairs of bonds is $s_n$,
while the corresponding shear critical stretch
is $s_t$, therefore the critical values of the cohesive energy
for each bond can be found as follows,
\begin{eqnarray}
\phi_{Ic}  &=&
{\phi_n} \left \{
1 +
\displaystyle
\exp \bigl(
- {\frac{s_n}{c_n}
} \bigr)
\left\{
(
\displaystyle
1-r+ { \frac{s_n}{c_n}}
)
{\frac{1-q}{r-1}}
-\bigl[
\displaystyle
q+(\frac{r-q}{r-1})
{ \frac{s_n}{c_n} }
\bigr]
\right\} \right \}~
\nonumber
\\
&=&\frac{1}{2}\phi_n\left[1-
\Bigl(1 + {\lambda_n^{-1}} \Bigr)
\exp{(-\lambda_n)} \right]
\label{eq:characteristic-length-4}
\\
\phi_{IIc} &=&
{\phi_n} \left \{
1 +
\displaystyle
\left\{
(
\displaystyle
1-r
)
{\frac{1-q}{r-1}}
-
\displaystyle
q
\exp \bigl(
-{s_t^2 \over c_t^2}
\bigr)
\right\} \right \}
\nonumber
\\
&=&
\frac{1}{2}q\phi_n\left[1-\exp{(-\lambda_t^2)}\right]
\label{eq:characteristic-length-5}
\end{eqnarray}
where we define
\begin{equation}
s_n := \lambda_n c_m~~~{\rm and}~~~~s_t := \lambda_t c_t
\end{equation}
which are amplitude factors for critical stretches.
\begin{figure}[H]
	\centering
	\includegraphics[height=2.5in]{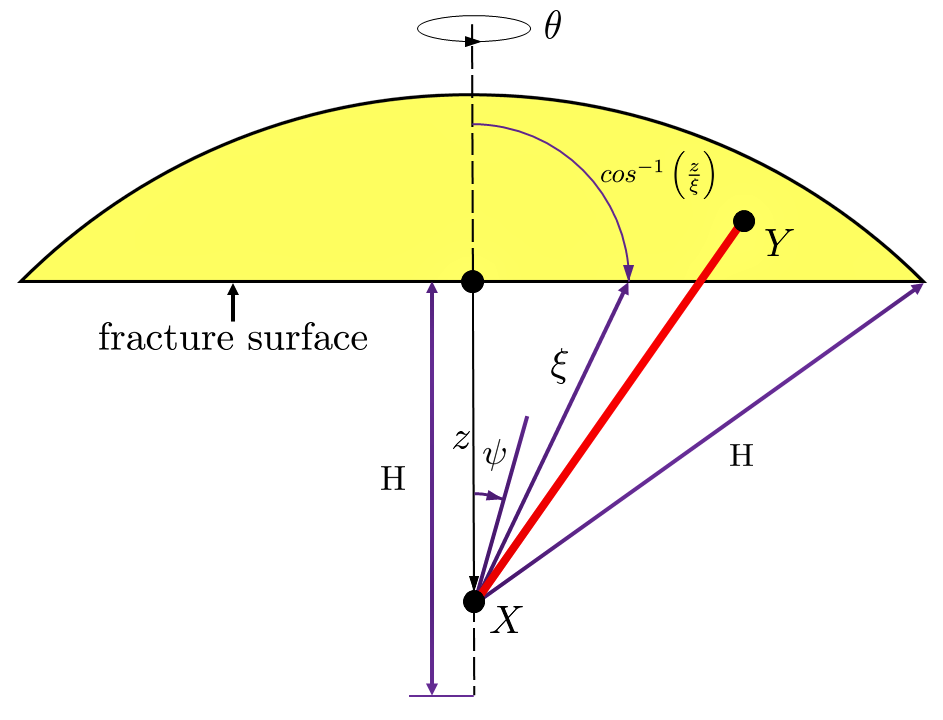}
	\caption{Integration domain for determine the critical energy release $G_0$.}
\label{fig:S0}
\end{figure}

Following \cite{Silling2005,Yu2020},
the corresponding energy releases can be related to the critical cohesive energy under
both tensile and shear deformation mode as follows,
\begin{eqnarray}
G_{I0} (\lambda_n) &=&  \int_0^H \int_0^{2 \pi} \int_z^H \int_0^{\cos^{-1}(z/\xi)}
\phi_{Ic} (\lambda_n)
\xi^2 \sin \psi d \psi d \xi d \theta d z
~
\label{eq:characteristic-length-4-1}
\\
G_{II0} (\lambda_t) &=&
\int_0^H \int_0^{2 \pi} \int_z^H \int_0^{\cos^{-1}(z/\xi)} \phi_{IIc} (\lambda_t)
\xi^2 \sin \psi d \psi d \xi d \theta d z
\label{eq:characteristic-length-5-1}
\end{eqnarray}
The above integrations are performed in an integration domain
shown in Fig. \ref{fig:S0}.
Based on the above equations,
we can find critical stretches $s_n$ and $s_t$ by
implicitly solving two nonlinear algebraic equations:
\begin{equation}
G_{Ic} - G_{I0} (\lambda_n) =0,~~{\rm and}~~ G_{IIc} - G_{II0} (\lambda_t) =0~.
\end{equation}

\begin{remark}
In the proposed cohesive peridynamics modeling (CPDM), the critical
stretch is not necessary concept. This is because that the functional form
of the mesoscale cohesive potential implicitly determines the critical 
stretch --- that is the critical stretch is a part of the cohesive potential,
and one does not need extra effort to break a bond.
All the material bond can be broken naturally without user interference.
However, the determination of $\lambda_s$ and $\lambda_t$ can help to set up
the cutoff distance. In this sense, the cutoff distance of the material bond should be chosen as
\[
\ell_c = s_c \ell_{\xi}  ,~~{\rm where}~~      s_c = min(\lambda_n c_n, \lambda_t c_t ).
\] 
\end{remark}

\section{Cohesive stress and peridynamic stress}
A main task of cohesive nonlocal continuum mechanics is
to find the underline cohesive continuum stress measures based
on the bond deformation and its corresponding bond force.
This is the step of micro to macro transition, which bridges
the mesoscale description and the macroscale description.
This will help us to understand a host of physical phenomena
from different perspectives, such as the crack growth criterion \cite{Yu2020}.

To proceed, we first recall Eq. (\ref{eq:fbar}),
\begin{eqnarray}
\boldsymbol{P}(\boldsymbol{X}) &=&
\frac{\partial W(\boldsymbol{X})}{\partial \boldsymbol{F}}
=
\frac{1}{2}\int_{\mathcal{H}}
\Bigl[
\bar{\bf f} \otimes \boldsymbol{\xi}
\Bigr]
d V_{\boldsymbol{\xi}}~
\label{eq:general-11}
\end{eqnarray}

Consider the following peridynamic force sampling formula,
\begin{equation}
\bar{\bf f} ({\bf X}^{\prime}, {\bf X})
= \sum_{I=1}^{N}\sum_{J=1, J \not =I}^{N}{\bf t}_{IJ}
w ({\bf X}_I - {\bf X})
\delta (({\bf X}_J-{\bf X}_I) - ({\bf X}^{\prime}-{\bf X}))~.
\label{eq:Irving1}
\end{equation}
where ${\bf X}, {\bf X}^{\prime}, {\bf X}_I$ and ${\bf X}_J$
are material particles in the referential configuration, ${\bf t}_{IJ}$ is
the bond force (not the force state) acting on the particle ${\bf X}_I$ from the particle
${\bf X}_J$.

By substituting the force sampling expression in Eq. (\ref{eq:Irving1}) into
Eq. (\ref{eq:general-11}), we have
\begin{eqnarray}
{\bf P}({\bf X}) &=&
\frac{1}{2}\int_{\mathcal{H}}
\Bigl[
\bar{\bf f} \otimes \boldsymbol{\xi}
\Bigr]
d V_{{\xi}}
\nonumber
\\
&=& \frac{1}{2} \int_{\mathcal{H}}
\sum_{I=1}^{N} \sum_{J=1, J \not =I}^{N}
w ({\bf X}_I - {\bf X})
{\bf t}_{IJ} \otimes {\bfg \xi}
\delta (({\bf X}_J-{\bf X}_I) - ({\bf X}^{\prime}-{\bf X})) d V_{{\xi}}~.
\end{eqnarray}

For simplicity,
we may choose the radial step function as the sampling function, i.e.
\begin{equation}
w (r) = \left \{
\begin{array}{lcl}
\displaystyle {1 \over \Omega_X}, && r < \delta
\\
\\
0 ,  && {\rm otherwise}
\end{array}
\right .
\label{eq:RadialS}
\end{equation}
where $\Omega_X = \mathcal{H}_X$ and $ vol(\mathcal{H}_X) = (4/3) \pi H^3$, and $H$
is the radius of
the horizon.

Since ${\bf X}, {\bf X}_I \in \mathcal{H}_X$, $w({\bf X}_I - {\bf X}) =1 $.
We then have the mathematical expression of the cohesive first Piola-Kirchhoff stress,
\\
\begin{tcolorbox}
\begin{eqnarray}
{\bf P}_{coh} ({\bf X}) &=& \frac{1}{2\Omega_X} \int_{\mathcal{H}}
\sum_{I=1}^{N} \sum_{J=1, J \not =I}^{N} {\bf t}_{IJ}
 \otimes {\bfg \xi}
\delta ({\bfg \xi}_{IJ} -{\bfg \xi}) d V_{{\xi}}~
\nonumber
\\
&=&  \frac{1}{2\Omega_X}
\sum_{I=1}^{N} \sum_{J=1, J \not =I}^{N} {\bf t}_{IJ}
 \otimes {\bfg \xi}_{IJ}
 \label{eq:C-stress}
\end{eqnarray}
\end{tcolorbox}

\smallskip
In 2008, based on Noll's lemma \cite{Noll1955},
Lehoucq and Silling \cite{Lehoucq2008} proposed a peridynamic stress \cite{Lehoucq2008},
\begin{equation}
{\bf P}_{LS} ({\bf X}) :=  {1 \over 2} \int_{\mathcal{S}^2}
\int_{0}^{\infty}
\int_{0}^{\infty} (y+z)^2
 {\bf f}({\bf X} +y {\bf M}, {\bf X} - z {\bf M}) \otimes
 {\bf M} dz d y d \Omega_M~,
\label{eq:Nonlocal-ST1}
\end{equation}
where $\mathcal{S}^2$ is the unit sphere.

Now, we show that the cohesive stress derived in Eq. (\ref{eq:C-stress})
is exactly the same as the peridynamic stress defined by Lehoucq and Silling \cite{Lehoucq2008},
i.e.
\[
{\bf P}_{coh} = {\bf P}_{LS}~.
\]

\begin{figure}[H]
\centering
\includegraphics[width=4.5in]{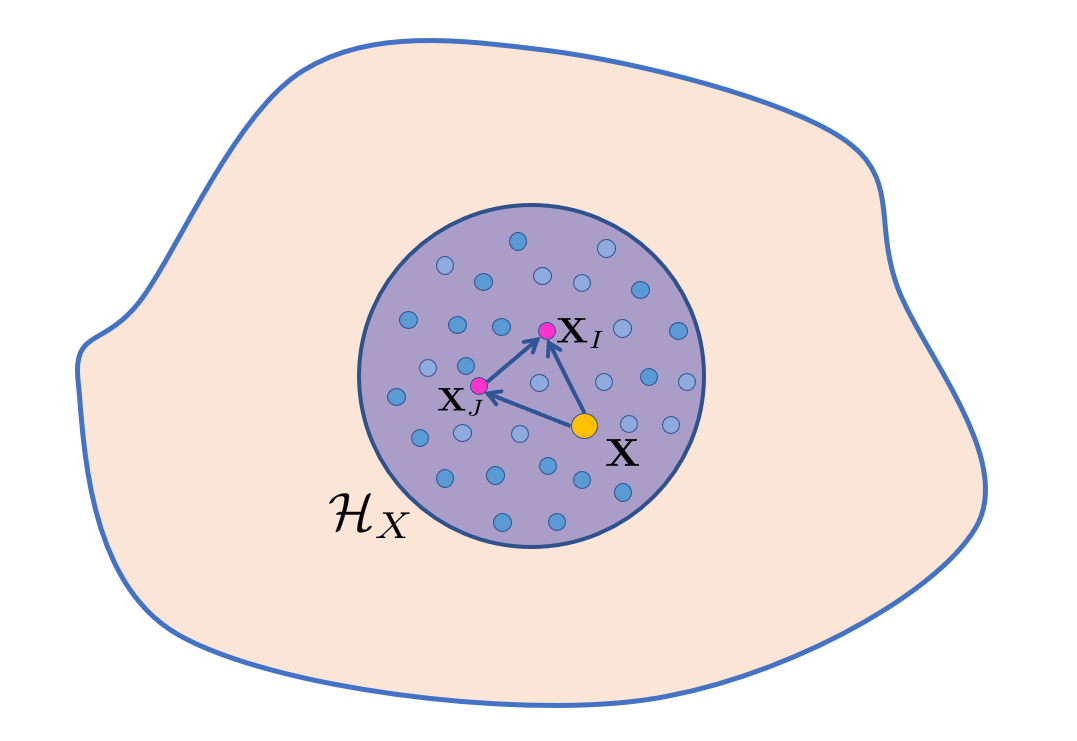}
\caption{
Illustration of peridynamics particle sampling strategy
}
\label{fig:fig1}
\end{figure}

\begin{theorem}[Peridynamic Stress]
~~{}\\
\smallskip
Assume that the average Peridynamic force density in a horizon
that can be expressed as the following discrete sampling expression of
the Irving-Kirkwood-Hardy formulation
\cite{Silling2010,Irving1950,Hardy1982},
\begin{equation}
\bar{\bf f}({\bf X}, {\bf X}^{\prime})
=
\sum_{I=1}^{N_X} \sum_{J=1, J\not =I}^{N_X} {\bf t}_{IJ} w ({\bf X}_I - {\bf X})
 \delta (({\bf X}_J -{\bf X}_I) - ({\bf X}^{\prime}-{\bf X})),
\end{equation}
where
${\bf X}, {\bf X}^{\prime}, {\bf X}_I$ and $ {\bf X}_J$ are
material particles in the referential configuration,
${\bf t}_{IJ}$
is the force (not the force state) acting
on the particle ${\bf X}_I$ from the particle ${\bf X}_J$ (see Fig. \ref{fig:fig1}),
$N_X$ is the total number of particles inside the horizon $\mathcal{H}_{X}$,
 $\delta ({\bf X})$
is the Dirac delta function,
and $w({\bf X}_I - {\bf X})$
is a window function or kernel function.

The nonlocal peridynamic stress defined by Lehoucq and Silling \cite{Lehoucq2008}
\begin{equation}
{\bf P}_{LS} :=  {1 \over 2} \int_{\mathcal{S}^2}
\int_{0}^{\infty}
\int_{0}^{\infty} (y+z)^2
 {\bf f}({\bf X} +y {\bf M}, {\bf X} - z {\bf M}) \otimes
 {\bf M} dz d y d \Omega_M~,
\label{eq:Nonlocal-BLM2}
\end{equation}
can be expressed the following discrete summation form,
\begin{equation}
{\bf P}_{LS} ({\bf X}) :=
{1 \over 2} \sum_{I=1}^{N_X} \sum_{J =1,J\not =I}^{N_X} {\bf t}_{IJ} \otimes
({\bf X}_J - {\bf X}_I) B_{IJ} ({\bf X}),~~{\bf X}_I, {\bf X}_J \in
\mathcal{H}_X,~~
\label{eq:Nonlocal-ST2}
\end{equation}
where ${\bf X}$ is the center point
of the horizon $\mathcal{H}_{X}$ and ${\bf X} \in \mathcal{B}$,
${\bf t}_{IJ} = {\bf f}({\bf X}_J, {\bf X}_I)V_I V_J$
is the force acting on the particle ${\bf X}_I$ by the particle ${\bf X}_J$,
where ${\bf X}_I, {\bf X}_J \in \mathcal{H}_{X}$, $V_I$ and $V_J$
represents the volume of material particle ${\bf X}_I$ and ${\bf X}_J$,
respectively, as shown in Fig. \ref{fig:fig1},
and
\begin{equation}
B_{IJ} ({\bf X}) = \int_0^1 w (\alpha ({\bf X}_{J} -{\bf X}_I) + {\bf X}_I
- {\bf X}) d \alpha
\end{equation}
is the bond function.
\end{theorem}

\begin{proof}
Based on Noll's lemma \cite{Noll1955}, we can write the
first peridynamics Piola-Kirchhoff stress as
\begin{eqnarray}
{\bf P}_{LS} ({\bf X}) &=& {1 \over 2 } \int_{\mathcal{S}^2} d \Omega_m
\int_0^{\infty} R^2 d R \int_0^1 {\bf f} ({\bf X} + \alpha R {\bf M},
{\bf X} - (1-\alpha) R {\bf M}) \otimes {\bf M} d \alpha
\nonumber
\\
&=&- {1 \over 2 } \int_{3} d V_R
\int_0^1 {\bf f} ({\bf X} + \alpha  {\bf R},
{\bf X} - (1-\alpha)  {\bf R}) \otimes {\bf R} d \alpha~,
~\forall {\bf X} \in \mathcal{B}
\label{eq:Ntilde2}
\end{eqnarray}

Considering the Hardy-Murdoch procedure
\cite{Hardy1982,Murdoch1983,Murdoch2007},
we have the following peridynamics sampling formulation (see Fig. 1)
\begin{equation}
{\bf f} ({\bf X}^{\prime}, {\bf X})
= \sum_{I=1}^{N_X}\sum_{J=1, J \not =I}^{N_X}{\bf t}_{IJ}
w ({\bf X}_I - {\bf X})
\delta (({\bf X}_J-{\bf X}_I) - ({\bf X}^{\prime}-{\bf X})),~~
\label{eq:Irving2}
\end{equation}
where the window function, or sampling function,
must satisfy the following conditions,
\begin{equation}
\int_{\mathcal{H}_X} w ({\bf y} - {\bf x}) d V_y = 1~,
\label{eq:cond1}
\end{equation}
and
\begin{equation}
\lim_{r \to 0} w(r) ~\to~\delta (r)~.
\label{eq:cond2}
\end{equation}

Condition (\ref{eq:cond1}) is the averaging requirement,
and Condition (\ref{eq:cond2}) ensures
that the Dirac comb sampling can converge to
a correct continuum form of integrand in Eq. (3), i.e.
\[
\sum_{I=1}^{N_X}\sum_{J=1, J \not =I}^{N_X}{\bf t}_{IJ}
w ({\bf X}_I - {\bf X})
\delta (({\bf X}_J-{\bf X}_I) - ({\bf X}^{\prime}-{\bf X}))
~\to~ {\bf f} ({\bf X}, {\bf X}^{\prime} - {\bf X})~.
\]

Letting
\[
{\bf X} = {\bf X} + \alpha {\bf R},
~{\rm and}~~ {\bf X}^{\prime} = {\bf X} -(1-\alpha) {\bf R}
\]
and substituting them into Eq. (\ref{eq:Irving2}), we then have
\begin{eqnarray}
&&{\bf f} ({\bf X} +\alpha {\bf R}, {\bf X} - (1-\alpha) {\bf R})
\nonumber
\\
&=& \sum_{I=1}^{N_X}\sum_{J=1, J\not =I}^{N_X} {\bf t}_{IJ}
w (({\bf X}_I - {\bf X})- \alpha {\bf R})
\delta ({\bf R} - ({\bf X}_I -{\bf X}_J)),
\label{eq:Irving3}
\end{eqnarray}
where ${\bf X}_I, {\bf X}_J \in \mathcal{H}_{X_C},~~{\bf X}_I \not = {\bf X}_J$.

Considering the following integration identities
\begin{eqnarray}
&&\int_{-\infty}^{\infty} \delta (\xi - x) w (x-\eta) d x = w(\xi -\eta),
\end{eqnarray}
we first integrate
\begin{eqnarray}
&&\int^{3}
\delta ({\bf R}- ({\bf X}_I - {\bf X}_J))
w (({\bf X}_I-{\bf X})) - \alpha {\bf R}) {\bf R} d  V_R
\nonumber
\\
&&=  ({\bf X}_I - {\bf X}_J )
w \bigl(({\bf X}_I-{\bf X}) - \alpha ({\bf X}_I - {\bf X}_J )\bigr)~.
\end{eqnarray}

Following \cite{Hardy1982}, we may define the second integral as
the so-called bond function, i.e.
 \begin{equation}
 B_{IJ} ({\bf X}) = \int_0^{1}
  w (\alpha({\bf X}_J-{\bf X}_I) + {\bf X}_I - {\bf X}) d \alpha
 \end{equation}
Thus, we have
\begin{eqnarray}
{\bf P}_{LS} ({\bf X}) &=& {1 \over 2 }
\bigl(
\sum_{I=1}^{N_X}\sum_{J=1, J \not = I}^{N_X} {\bf t}_{IJ}
 \otimes ({\bf X}_J - {\bf X}_I) \bigr) B_{IJ} ({\bf X})~,
\label{eq:Ntilde3}
\end{eqnarray}
which is called the Hardy stress (see \cite{Hardy1982,Zimmerman2004}.
\begin{figure}[H]
\centering
\includegraphics[width=4.0in]{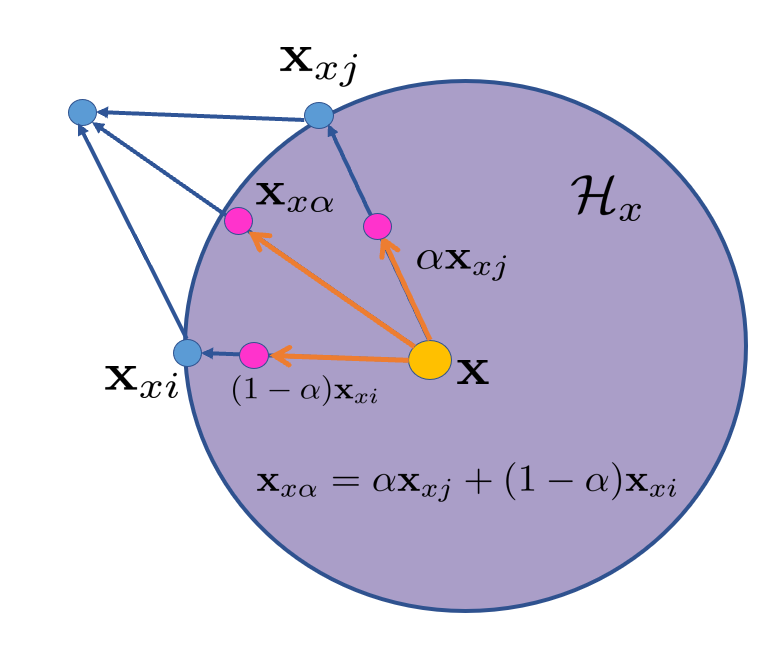}
\caption{
Graphic illustration of
the bond integration variable ${\bf X}_{x\alpha} \in \mathcal{H}_X$,
where ${\bf X}_{x\alpha} = \alpha {\bf X}_{xJ} + (1-\alpha) {\bf X}_{xI}$.
}
\label{fig:fig6}
\end{figure}

If we choose $w({\bf x})$ as the spherical radial step function
(see Eq. (\ref{eq:RadialS})), one can see that
\[
w(\alpha({\bf X}_J - {\bf X}_I) + {\bf X}_I -{\bf X})
=w(\alpha({\bf X}_J-{\bf X})+ (1-\alpha) ({\bf X}_I - {\bf X}))~.
\]
If ${\bf X}_I, {\bf X}_J \in \mathcal{H}_X$, we can see that
\[
{\bf X}_{x\alpha}:= \alpha({\bf X}_J-{\bf X})+ (1-\alpha) ({\bf X}_I - {\bf X}) \in \mathcal{H}_X
\]
This is because that
\[
|\alpha({\bf X}_J-{\bf X})+ (1-\alpha) ({\bf X}_I - {\bf X})| \leq |\alpha({\bf X}_J-{\bf X})
+ (1-\alpha) ({\bf X}_J - {\bf X})| = |{\bf X}_J - {\bf X}| \le \delta
\]
if $|{\bf X}_J - {\bf X}| \ge |{\bf X}_I - {\bf X}|$, and vice vera
\[
|\alpha({\bf X}_J-{\bf X})+ (1-\alpha) ({\bf X}_I - {\bf X})| \leq |\alpha({\bf X}_I-{\bf X})
+ (1-\alpha) ({\bf X}_I - {\bf X})| = |{\bf X}_I - {\bf X}| \le \delta
\]
if $|{\bf X}_J - {\bf X}| \le |{\bf X}_I - {\bf X}|$ as shown in Fig. \ref{fig:fig6}.

Thus, it is readily to show that
\[
B_{IJ} ({\bf X}) = {1 \over \Omega_X},~~~~{\rm if}~{\bf X}_I, {\bf X}_J \in \mathcal{H}_X
\]

For this special case, the peridynamic stress has the expression,
\begin{tcolorbox}
\begin{eqnarray}
{\bf P}_{LS} ({\bf X}) &=& {1 \over 2 \Omega_X }
\bigl(
\sum_{I=1}^{N}\sum_{J=1, J \not = I}^{N} {\bf t}_{IJ}
 \otimes ({\bf X}_J - {\bf X}_I) \bigr) ~.
\label{eq:Ntilde4}
\end{eqnarray}
\end{tcolorbox}
Equation (\ref{eq:Ntilde4}) confirms
that the peridynamic stress is
the first Piola-Kirchhoff virial stress, or it is equal to the cohesive
first Piola-Kirchhoff stress.
\end{proof}

\section{Numerical examples}
In this section,
we present several numerical examples to validate the proposed CPDM method.
All the 2D models or examples are computed under 2D plane stress conditions
using uniform particles.
For the 3D example,
we conducted a three-point bending beam test,
which is widely used to investigate mix-mode fracture behavior.
Force-displacement curves for all cases are compared with experimental results.

\subsection{Two-dimensional crack growth}
\begin{figure}[H]
	\begin{center}
		\includegraphics[height=3.0in]{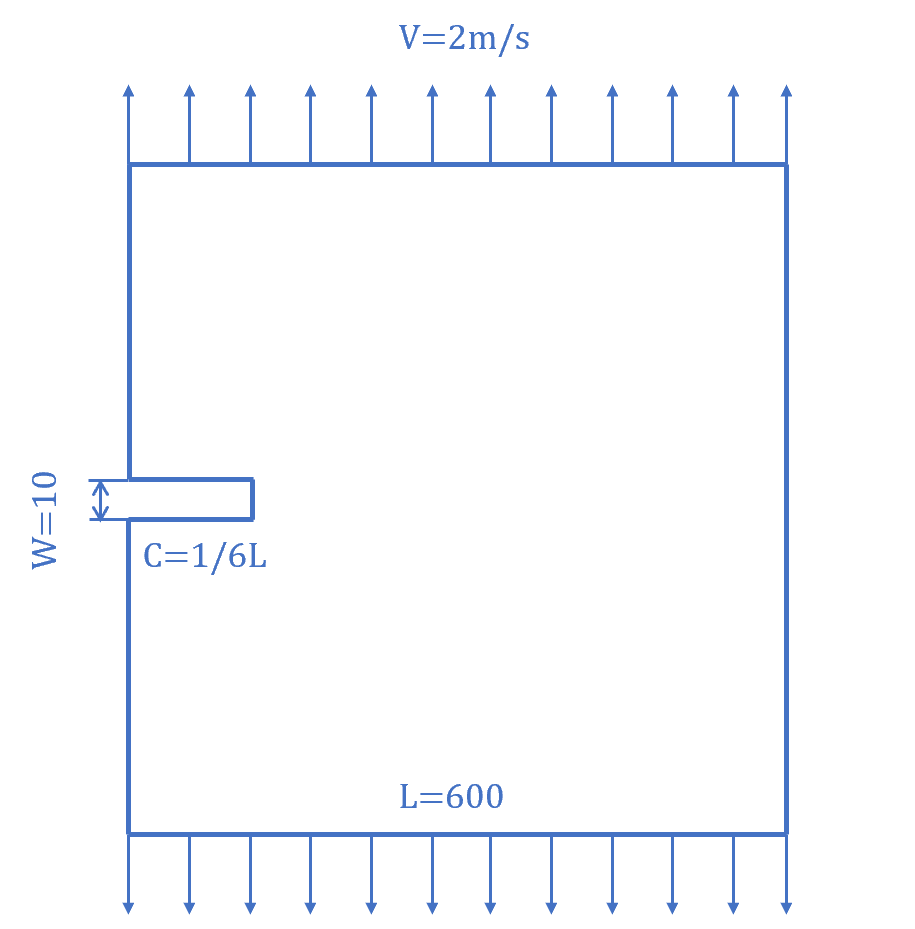}
	\end{center}
	\caption{Sketch map of 2D crack test}
	\label{fig:cracksetting}
\end{figure}

\begin{figure}[H]	
	\centering
	\begin{subfigure}[t]{0.24\linewidth}
		\includegraphics[height=1.5in]{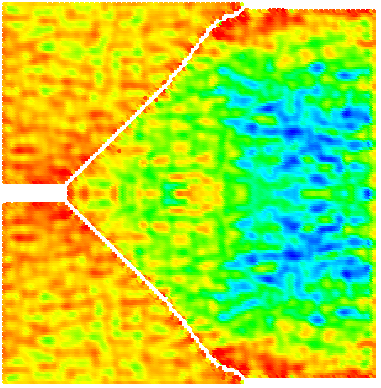}
		\caption{$\delta_n=\delta_t$}\label{fig:deltat0.5}		
	\end{subfigure}
	\begin{subfigure}[t]{0.24\linewidth}
		\centering
		\includegraphics[height=1.5in]{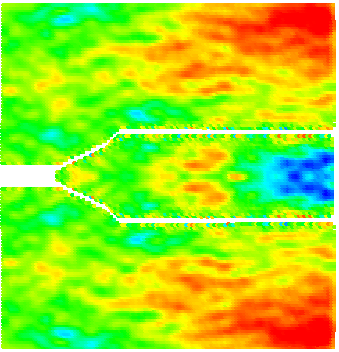}
		\caption{$\delta_n=\delta_t$}\label{fig:deltat1}
	\end{subfigure}
	\begin{subfigure}[t]{0.24\linewidth}
		\centering
		\includegraphics[height=1.5in]{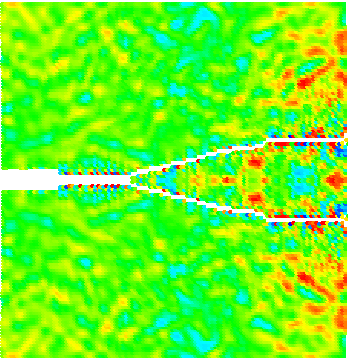}
		\caption{$\delta_n=5\delta_t$}\label{fig:deltat5}
	\end{subfigure}
	\begin{subfigure}[t]{0.03\linewidth}
		\centering
		\includegraphics[height=1.5in,width=0.4in]{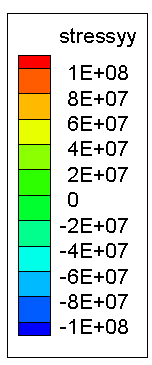}
	\end{subfigure}
	\hfill
	\centering
	\begin{subfigure}[t]{0.25\linewidth}
		\centering
		\includegraphics[height=1.5in]{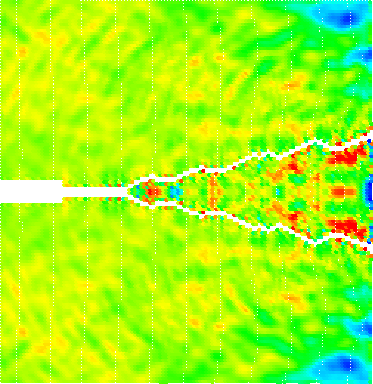}
		\caption{$\delta_n=10\delta_t$}\label{fig:deltat10}
	\end{subfigure}
	\begin{subfigure}[t]{0.25\linewidth}
		\centering
		\includegraphics[height=1.5in]{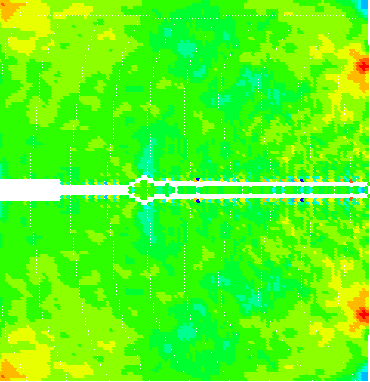}
		\caption{$\delta_n=20\delta_t$}\label{fig:deltat20}
	\end{subfigure}
	\begin{subfigure}[t]{0.04\linewidth}
		\centering
		\includegraphics[height=1.5in,width=0.5in]{crackstress.png}
	\end{subfigure}
	\caption{Crack patterns with respect to the different ratios of $\delta_n/\delta_t$}
\label{fig:crackdeltat}
\end{figure}

In this example, we used CPDM to simulate a 2D crack growth problem
to validate the proposed CPDM method.
The specimen size and boundary condition setting are shown
 in Fig.\ref{fig:cracksetting}.
 The morphology of the specimen after complete fracture is shown as the result diagram in Fig.
 \ref{fig:crackdeltat}, and the color contour represents $S_{22}$ distribution
 (PK-II stress component in y-direction).

An advantage of using cohesive mesoscale potential
is that by adjusting the numerical values of the parameters we can observe
both brittle and ductile fracture as well as their transition .
For example, by adjusting the ratio of the parameter $\delta_n$ to $\delta_t$,
the simulated crack shape changes from ductile fracture to brittle fracture
as shown in Fig. \ref{fig:crackdeltat}.

\begin{figure}[H]	
	\centering
	\begin{subfigure}[t]{0.3\linewidth}
		\centering
		\includegraphics[height=1.65in]{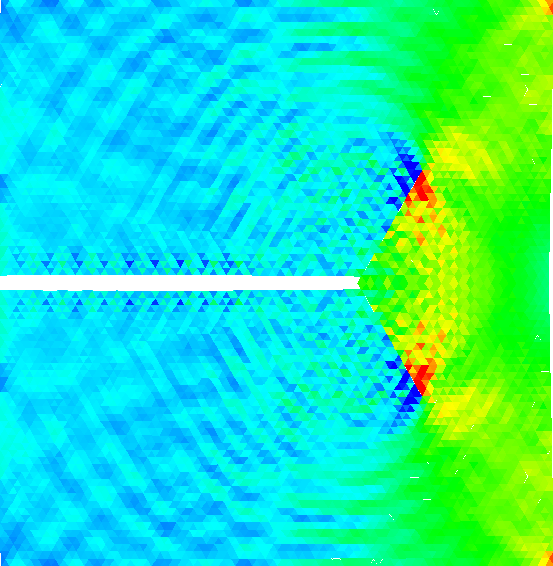}
		\caption{Results of FEM-CZM}\label{fig:femczm1}		
	\end{subfigure}
	\begin{subfigure}[t]{0.3\linewidth}
		\centering
		\includegraphics[height=1.65in]{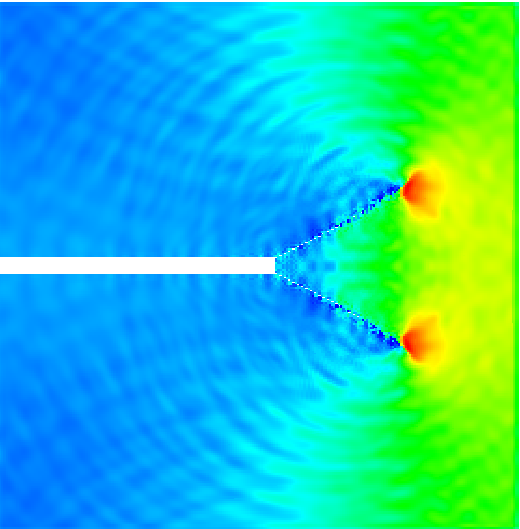}
		\caption{Results of CPDM}\label{fig:pdczm1}
	\end{subfigure}
    \begin{subfigure}[t]{0.05\linewidth}
    	\centering
    	\includegraphics[height=1.67in,width=0.65in]{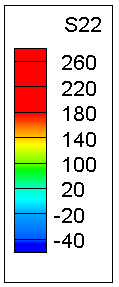}
    \end{subfigure}

    \caption{Comparison of FEM-CZM simulation with CPDM simulation in 2D crack propagation.}
    \label{fig:Comparisonfem}
\end{figure}

\begin{figure}[H]	
	\centering
	\begin{subfigure}[t]{0.23\linewidth}
		\centering
		\includegraphics[height=1.4in,width=1.4in]{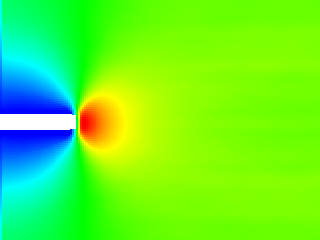}
		\caption{t=0.176ms}\label{fig:crack5}		
	\end{subfigure}
	\begin{subfigure}[t]{0.23\linewidth}
		\centering
		\includegraphics[height=1.4in,width=1.4in]{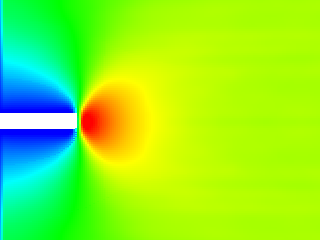}
		\caption{t=0.192ms}\label{fig:crack6}
	\end{subfigure}
    \begin{subfigure}[t]{0.23\linewidth}
    	\centering
    	\includegraphics[height=1.4in,width=1.4in]{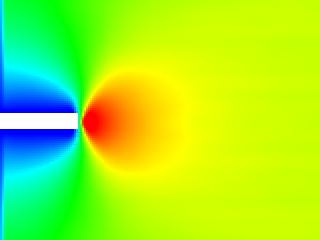}
    	\caption{t=0.242ms}\label{fig:crack7}
    \end{subfigure}
	\begin{subfigure}[t]{0.04\linewidth}
		\centering
		\includegraphics[height=1.4in,width=0.4in]{crackstress.png}
	\end{subfigure}

    \begin{subfigure}[t]{0.23\linewidth}
    	\centering
    	\includegraphics[height=1.4in,width=1.4in]{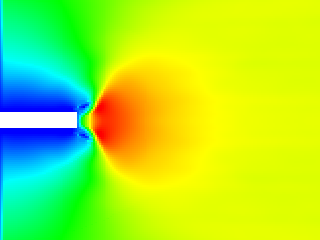}
    	\caption{t=0.253ms}\label{fig:crack8}		
    \end{subfigure}
    \begin{subfigure}[t]{0.23\linewidth}
    	\centering
    	\includegraphics[height=1.4in,width=1.4in]{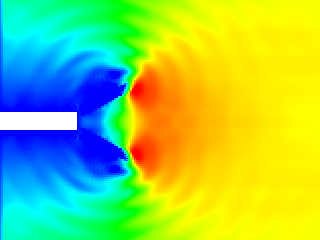}
    	\caption{t=0.258ms}\label{fig:crack9}
    \end{subfigure}
    \begin{subfigure}[t]{0.23\linewidth}
    	\centering
    	\includegraphics[height=1.4in,width=1.4in]{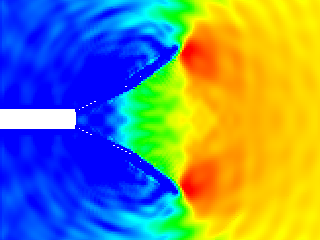}
    	\caption{t=0.269ms}\label{fig:crack10}
    \end{subfigure}
    \begin{subfigure}[t]{0.04\linewidth}
    	\centering
    	\includegraphics[height=1.4in,width=0.4in]{crackstress.png}
    \end{subfigure}
	\caption{Stress distribution at crack tip at several typical moments}
	\label{fig:crackstress}
\end{figure}

It can be seen from Fig.\ref{fig:crackdeltat} that from $(a)$ to $(e)$
with the increase of the ratio of $\delta_n/\delta_t$,
the feature of the brittle fracture gradually becomes obvious.
We also compared the calculation results of CPDM with those of FEM-CZM,
which are shown in Fig. \ref{fig:Comparisonfem}.
The ratio of $\delta_n/\delta_t$ in the comparison example is chosen as 0.2.
From \ref{fig:Comparisonfem}, one may find that
that the two results are in a good agreement.
In addition, we also simulated the transit effect of the stress evolution at the crack tip
during the crack propagation by using the notched specimen.
Figure \ref{fig:crackstress} shows the stress distribution of $S_{22}$
at the crack tip from the moment of crack initiation to the stage that crack propagated well into
the middle .
In this case, the ratio of $\delta_n/\delta_n$ is chosen as 1.

\subsection{Trunk's test: Wedge splitting fracture}
In 1999, Trunk \cite{2000Einfluss} conducted an experimental investigation
into the size dependence of non-linear fracture mechanics parameters for cementitious materials.
The wedge splitting test is adopted in his research,
which was later widely replicated for the purpose of verifying and validating various fracture simulations.
In this work, we also conducted a numerical experiment of Trunk's test
by using the proposed CPDM method to simulate wedge splitting fracture test.

\begin{figure}[H]
	\begin{center}
		\includegraphics[height=3.2in]{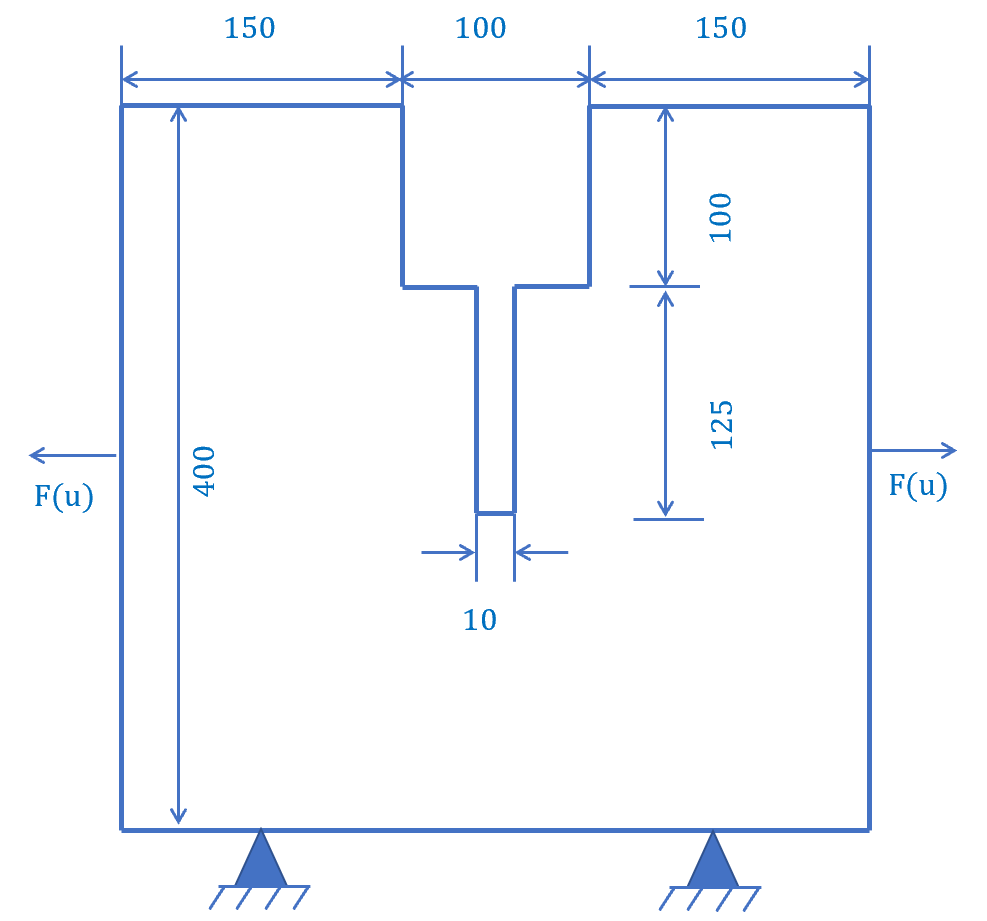}
	\end{center}
	\caption{Schematic illustration of wedge splitting test.}
	\label{fig:wedge experiment-setup}
\end{figure}

The geometry and boundary conditions are depicted in Fig. \ref{fig:wedge experiment-setup},
and the dimensions of the square-shaped specimen is chosen as 400mm$\times$400mm$\times$400mm, while
the width of the prefabricated crack is $10mm$.
Other geometric parameters are shown in Fig.\ref{fig:wedge experiment-setup}.
The material properties of the specimen adopted in the simulation are given as follows:
Young' modules $E=28.3Gpa$, Poisson's ratio $\nu=0.2$, and the fracture energy release $G_f=0.017n/mm$.
The specimen is subjected to a prescribed force at left and right side,
while the bottom of the specimen is fixed. The particle spacing is $\Delta=5mm$,
and the horizon radius $\delta$ is equal to 3$\Delta$.

\begin{figure}[H]	
	\centering
	\begin{subfigure}[t]{0.25\linewidth}
		\centering
		\includegraphics[height=1.4in]{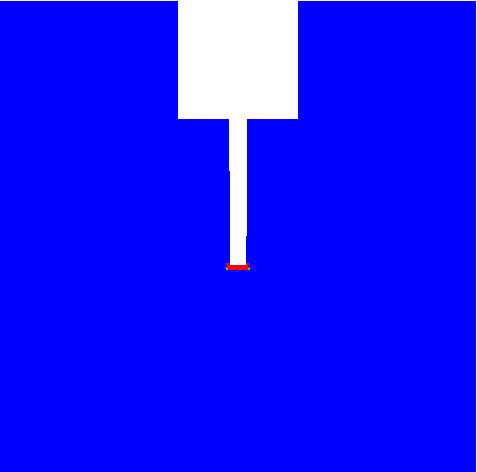}
		\caption{t=0ms}\label{fig:1dmg25}		
	\end{subfigure}
	\begin{subfigure}[t]{0.25\linewidth}
		\centering
		\includegraphics[height=1.4in]{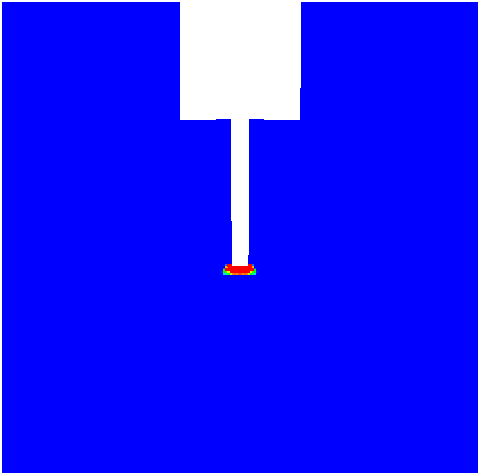}
		\caption{t=0.5ms}\label{fig:1dmg50}
	\end{subfigure}
	\begin{subfigure}[t]{0.25\linewidth}
		\centering
		\includegraphics[height=1.4in]{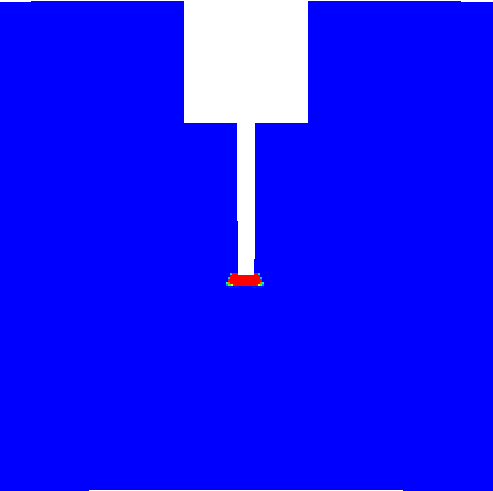}
		\caption{t=1.0ms}\label{fig:1dmg75}
	\end{subfigure}
	\begin{subfigure}[t]{0.05\linewidth}
		\centering
		\includegraphics[height=1.3in,width=0.95in]{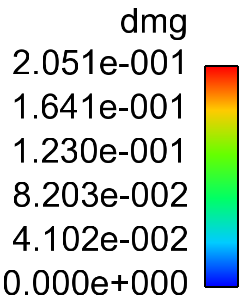}
	\end{subfigure}
	\hfill
	\begin{subfigure}[t]{0.25\linewidth}
		\centering
		\includegraphics[height=1.4in]{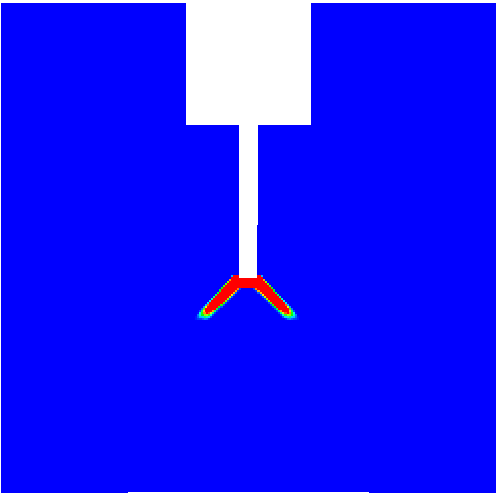}
		\caption{t=1.5ms}\label{fig:1dmg85}
	\end{subfigure}
	\begin{subfigure}[t]{0.25\linewidth}
		\centering
		\includegraphics[height=1.4in]{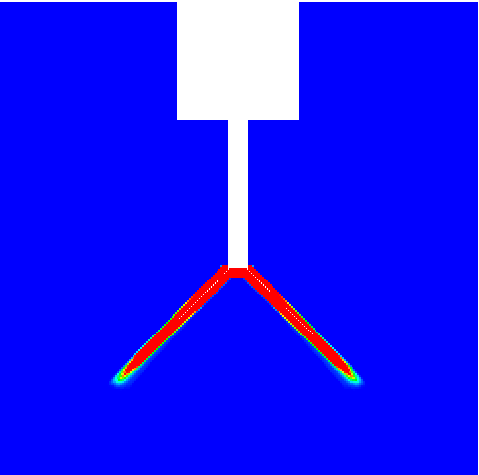}
		\caption{t=1.75ms}\label{fig:1dmg95}
	\end{subfigure}
	\begin{subfigure}[t]{0.25\linewidth}
		\centering
		\includegraphics[height=1.4in]{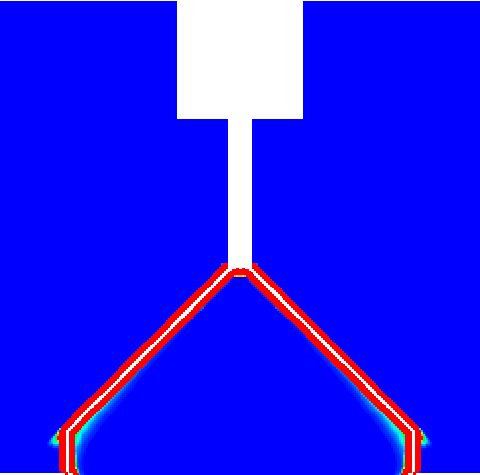}
		\caption{t=2.0ms}\label{fig:1dmg100}
	\end{subfigure}
	\begin{subfigure}[t]{0.05\linewidth}
		\centering
		\includegraphics[height=1.3in,width=0.95in]{wedge_color2.png}
	\end{subfigure}
	\caption{damage at several typical moments}\label{fig:wedgedmg}
\end{figure}

\begin{figure}[H]	
	\centering
	\begin{subfigure}[t]{0.25\linewidth}
		\centering
		\includegraphics[height=1.4in,width=1.45in]{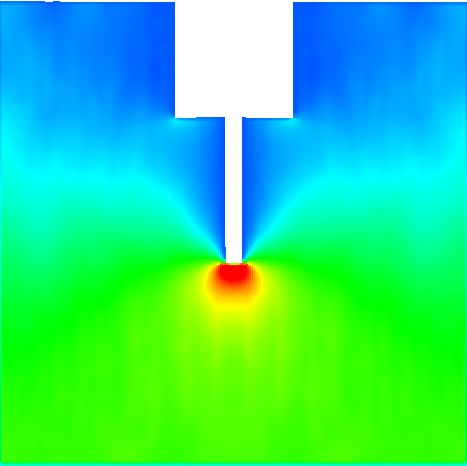}
		\caption{t=0ms}\label{fig:1stress1}		
	\end{subfigure}
	\begin{subfigure}[t]{0.25\linewidth}
		\centering
		\includegraphics[height=1.4in]{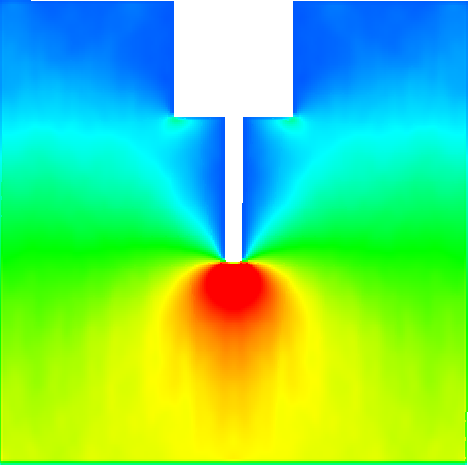}
		\caption{t=0.5ms}\label{fig:1stress25}
	\end{subfigure}
	\begin{subfigure}[t]{0.25\linewidth}
		\centering
		\includegraphics[height=1.4in]{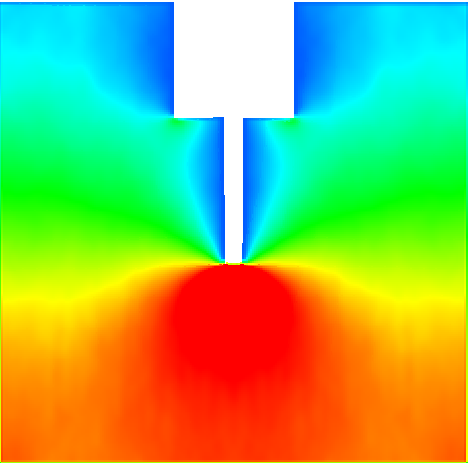}
		\caption{t=1.0ms}\label{fig:1stress50}
	\end{subfigure}
	\begin{subfigure}[t]{0.05\linewidth}
	\centering
	\includegraphics[height=1.35in,width=0.95in]{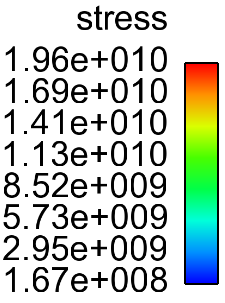}
    \end{subfigure}
	\hfill
	\begin{subfigure}[t]{0.25\linewidth}
		\centering
		\includegraphics[height=1.4in]{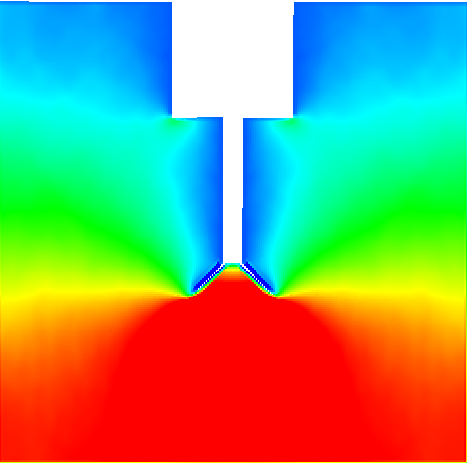}
		\caption{t=1.5ms}\label{fig:1stress75}
	\end{subfigure}
	\begin{subfigure}[t]{0.25\linewidth}
		\centering
		\includegraphics[height=1.4in]{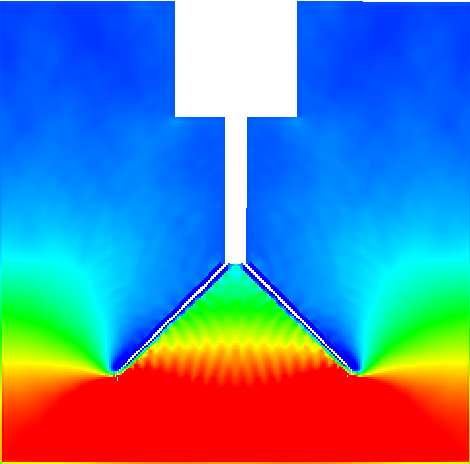}
		\caption{t=1.75ms}\label{fig:1stress85}
	\end{subfigure}
	\begin{subfigure}[t]{0.25\linewidth}
		\centering
		\includegraphics[height=1.4in]{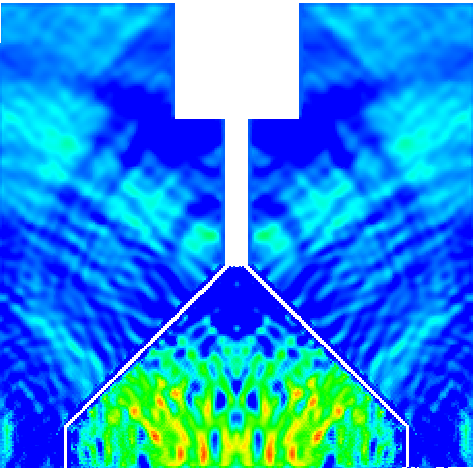}
		\caption{t=2.0ms}\label{fig:1stress100}
	\end{subfigure}
	\begin{subfigure}[t]{0.05\linewidth}
	\centering
	\includegraphics[height=1.35in,width=0.95in]{wedge_color.png}
    \end{subfigure}
	\caption{Stress distribution at several typical moments}
\label{fig:wedgestress}
\end{figure}

The simulated wedge crack splitting process is shown in Fig. \ref{fig:wedgedmg} with damage
color contour and in Fig. \ref{fig:wedgestress} with the stress color contour.
The sequences reflect the damage of the specimen and stress variables
at different time instances in the simulation.

Figure \ref{fig:wedgedmg} shows the change of damage contour
with respect to time.
Material damage is calculated for each material point.
In traditional PD method, for each material point, when the elongation of the bond between
the two material points
exceeds the critical stretch $s_0$, the irreversible fracture or bond breaking will occur.
We call the ratio of the number of broken bonds to the total number of bonds
of a material point its damage \cite{Jafarzadeh2019}, i.e.
\[
d ({\bf X}) = 1 -
{
\int_{\mathcal{H}_X} \mu ({\bf X}, {\bf X}^{\prime}) dV_{ {\bf X}^{\prime}}
\over
\int_{\mathcal{H}_X} d V_{{\bf X}^{\prime}}
}
\]
where
\[
 \mu ({\bf X}, {\bf X}^{\prime}) = \left \{
 \begin{array}{ll}
 1 &~{\rm if} ~\overline{{\bf X}{\bf X}^{\prime}}~ {\rm bond~is~broken}
 \\
 \\
 0 &~{\rm if}~\overline{{\bf X}{\bf X}^{\prime}}~ {\rm bond~is~not~broken}
 \end{array}
 \right .
\]
is the characteristic function of the material bond.

When the damage factor of a material point is 1,
it means that all bonds are not damaged.
When the damage factor is equal to 0, it means that the material point has been completely damaged.
As shown in Fig. \ref{fig:wedgedmg},
the wedge specimen began suffering damages in about 1ms,
and it was tore obviously at the middle prefabricated crack in about 1.5ms, and
the specimen was completely torn apart after 1.75ms.
The stress evolution process is shown in Fig.\ref{fig:wedgestress},
and the stress evolution process is basically consistent with that of the damage evolution.

\subsection{Fracture of L-shape plate}
In this example, to validate the proposed CPDM,
we apply it simulating the fracture of a L-shape concrete plate, which
was studied in \cite{2010Winkler} by using a finite element analysis of elasto-plastic
damage constitutive modeling.

\begin{figure}[H]
	\begin{center}
		\includegraphics[height=3.5in]{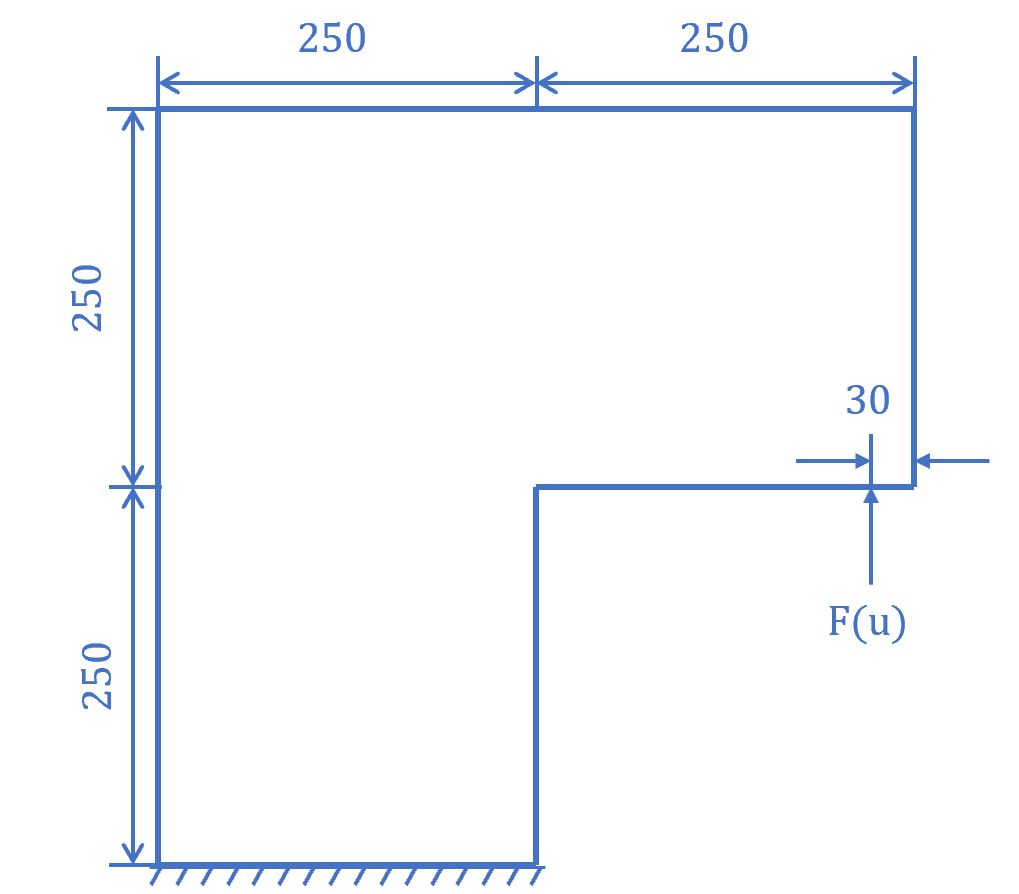}
	\end{center}
	\caption{Sketch map of L shape test}
	\label{fig:L shape experiment-setup}
\end{figure}

The geometry and boundary conditions of the L shape plate are depicted in
Fig. \ref{fig:L shape experiment-setup}.
As shown in the figure, L shape plate side length is 500mm,
the bottom end is a fixed boundary, and there is an upward force on the right bottom edge.
The material properties of the plate are chosen as follows:
$E=25.85 Gpa$ , $\mu$=$0.2$ and $G_{f}=0.015N/mm$.
In the numerical simulations, the spacing between particles is $5mm$.

\begin{figure}[H]	
	\centering
	\begin{subfigure}[t]{0.25\linewidth}
		\centering
		\includegraphics[height=1.4in]{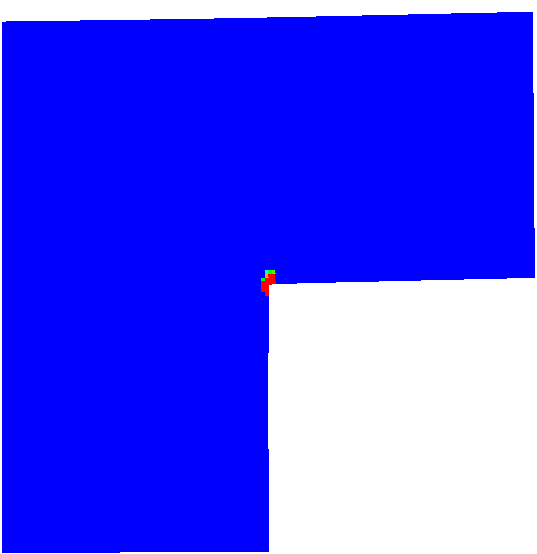}
		\caption{t=1.74ms}\label{fig:2dmg32}		
	\end{subfigure}
	\begin{subfigure}[t]{0.25\linewidth}
		\centering
		\includegraphics[height=1.4in]{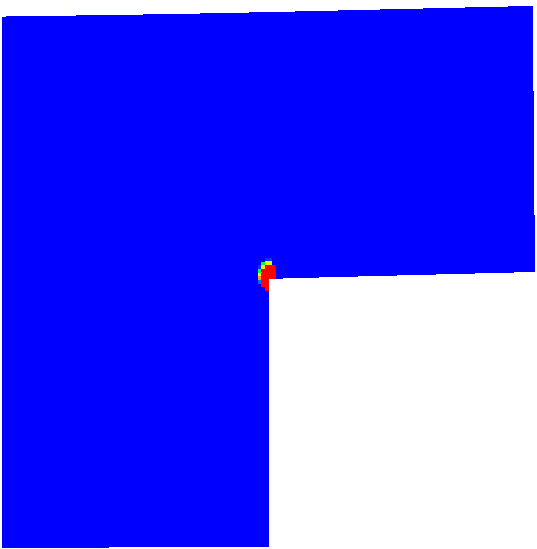}
		\caption{t=3.48ms}\label{fig:2dmg64}
	\end{subfigure}
	\begin{subfigure}[t]{0.25\linewidth}
		\centering
		\includegraphics[height=1.4in]{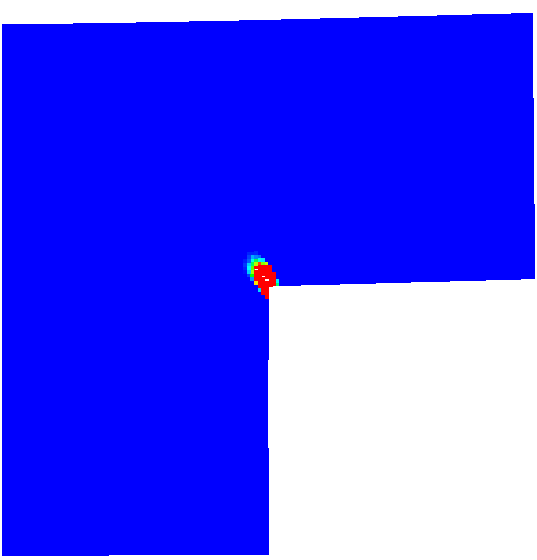}
		\caption{t=5.22ms}\label{fig:2dmg96}
	\end{subfigure}
	\begin{subfigure}[t]{0.05\linewidth}
	\centering
	\includegraphics[height=1.35in,width=0.9in]{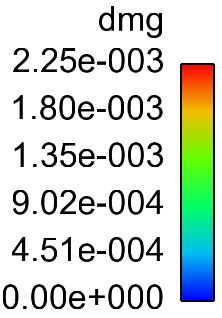}
    \end{subfigure}
	\hfill
	\begin{subfigure}[t]{0.25\linewidth}
		\centering
		\includegraphics[height=1.4in]{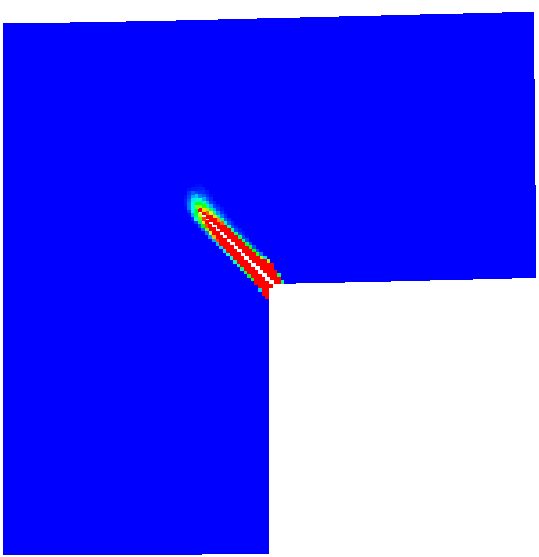}
		\caption{t=6.96ms}\label{fig:2dmg128}
	\end{subfigure}
	\begin{subfigure}[t]{0.25\linewidth}
		\centering
		\includegraphics[height=1.4in]{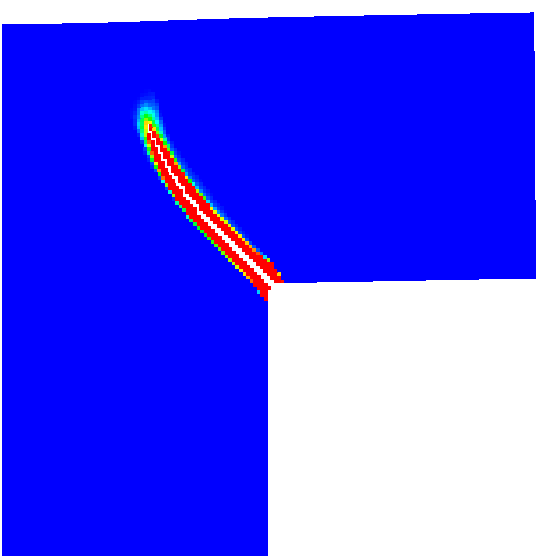}
		\caption{t=8.7ms}\label{fig:2dmg162}
	\end{subfigure}
    \begin{subfigure}[t]{0.25\linewidth}
    	\centering
    	\includegraphics[height=1.4in]{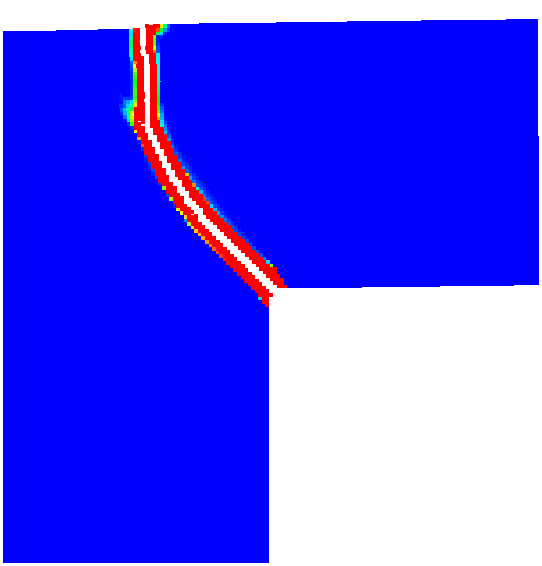}
    	\caption{t=10.4ms}\label{fig:2dmg163}
    \end{subfigure}
	\begin{subfigure}[t]{0.05\linewidth}
	\centering
	\includegraphics[height=1.35in,width=0.9in]{Lshape_color2.png}
    \end{subfigure}
	\caption{damage at several typical moments}\label{Lshapedmg}
\end{figure}

\begin{figure}[H]	
	\centering
	\begin{subfigure}[t]{0.25\linewidth}
		\centering
		\includegraphics[height=1.4in]{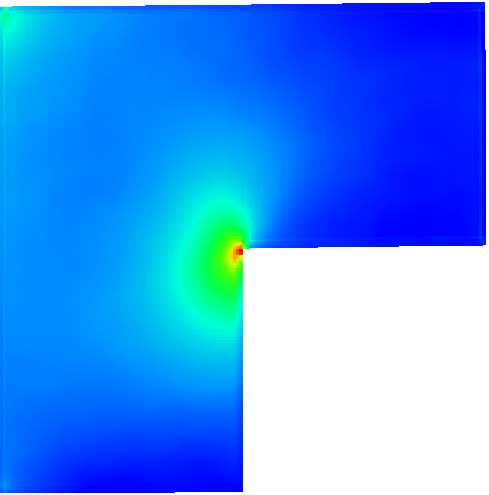}
		\caption{t=1.74ms}\label{fig:2stress32}		
	\end{subfigure}
	\begin{subfigure}[t]{0.25\linewidth}
		\centering
		\includegraphics[height=1.4in]{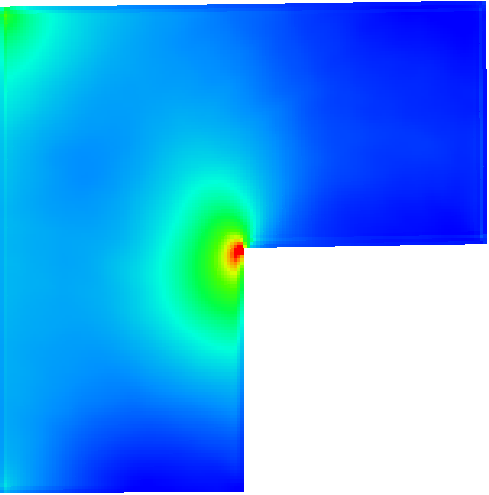}
		\caption{t=3.48ms}\label{fig:2stress64}
	\end{subfigure}
	\begin{subfigure}[t]{0.25\linewidth}
		\centering
		\includegraphics[height=1.4in]{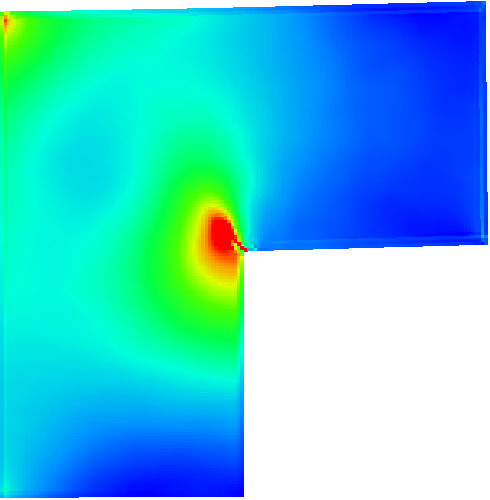}
		\caption{t=5.22ms}\label{fig:2stress96}
	\end{subfigure}
	\begin{subfigure}[t]{0.05\linewidth}
	\centering
	\includegraphics[height=1.35in,width=0.9in]{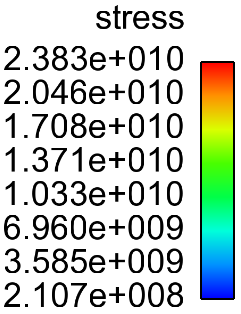}
    \end{subfigure}
	\hfill
	\begin{subfigure}[t]{0.25\linewidth}
		\centering
		\includegraphics[height=1.4in]{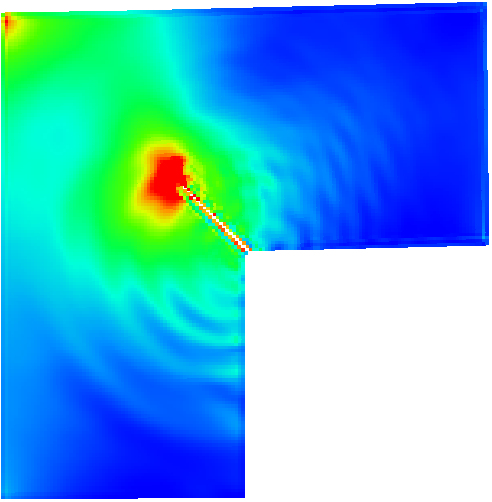}
		\caption{t=6.96ms}\label{fig:2stress129}
	\end{subfigure}
	\begin{subfigure}[t]{0.25\linewidth}
		\centering
		\includegraphics[height=1.4in]{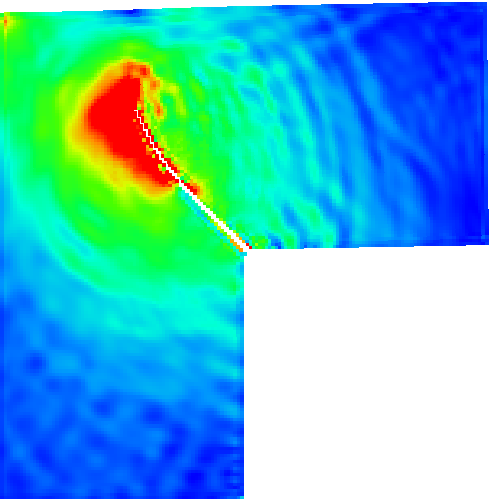}
		\caption{t=8.7ms}\label{fig:2stress162}
	\end{subfigure}
    \begin{subfigure}[t]{0.25\linewidth}
    	\centering
    	\includegraphics[height=1.4in]{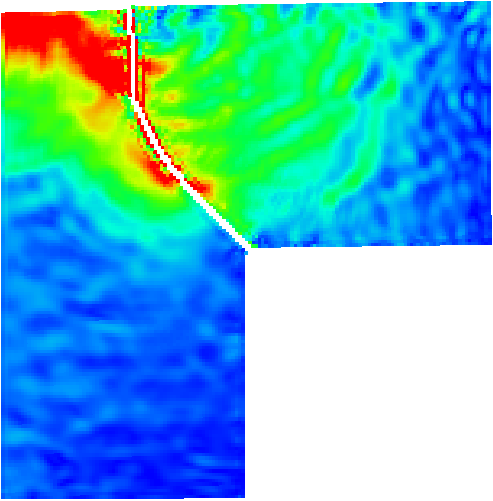}
    	\caption{t=10.4ms}\label{fig:2stress163}
    \end{subfigure}
	\begin{subfigure}[t]{0.05\linewidth}
	\centering
	\includegraphics[height=1.35in,width=0.9in]{Lshape_color.png}
    \end{subfigure}
	\caption{Stress distribution at several typical moments}
\label{Lshapestress}
\end{figure}

To validate the proposed CPDM formulation, the L-shape plate has
the same dimension and material constants used in \ref{Lshapedmg}, except
that we adopt the Xu-Needleman cohesive potential rather an elasto-plastic
damage model, while using nonlocal cohesive peridyanmics rather than
finite element crack smearing techniques.
Fig.\ref{Lshapestress} show the damage and stress distribution of the
L-shape plate.
It can be clearly seen from the resulting figure that at $5.22ms$,
the crack originated at the right angle and gradually extended to the left side,the L shape plate began
to suffer damage at the bend corner, and stress concentration also appeared at the bend corner
in Fig. \ref{Lshapestress}. The stress at the right margin in Fig. \ref{Lshapestress}
is obviously greater than the rest of the plate because this is where the external force is applied,
this is why this area also appear some damage.


\subsection{Plate with the hole}
In this example, we employed CPDM carrying out numerical
simulation of a thin plate with hole under uniaxial tension,
which is under plane stress condition.
For infinitesimal deformation, this problem has a close-form
solution, i.e. the well-known Kirsch solution \cite{Kirsch1898}.

\begin{figure}
	\centering
	\includegraphics[height=2.8in]{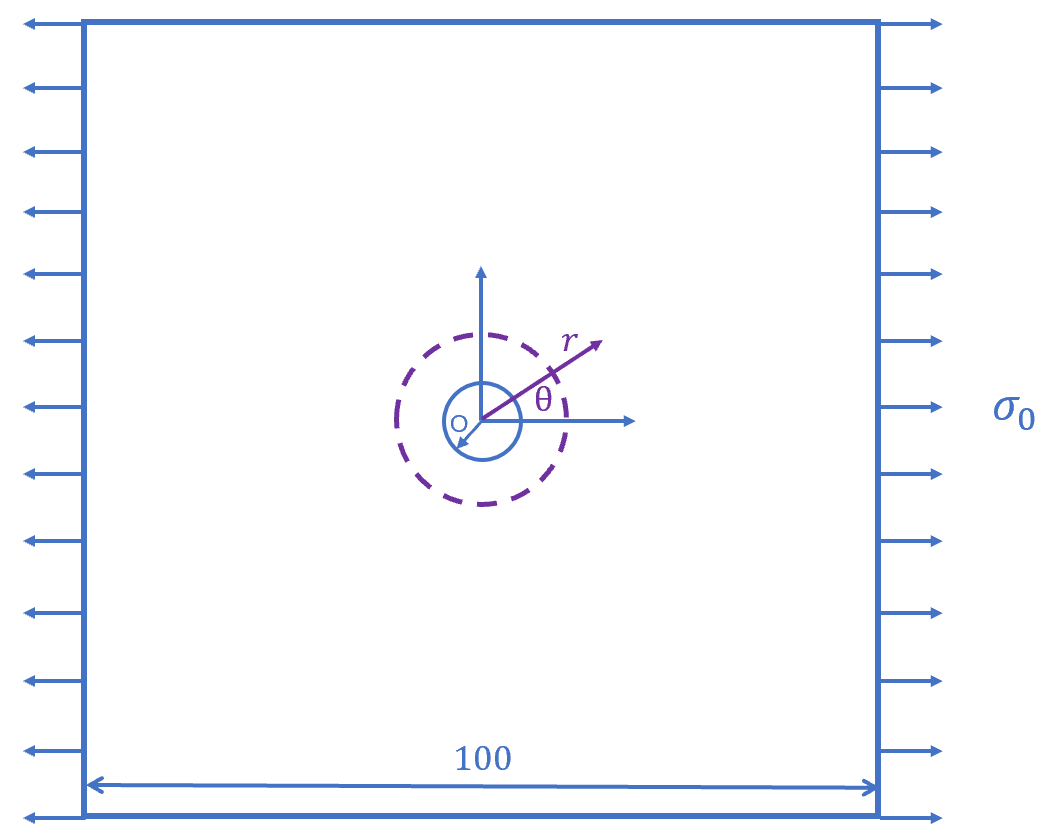}
	\caption{Schematic illustration of a plate with hole}
\label{fig:phsetting}
\end{figure}

\begin{figure}[H]
	\centering
	\includegraphics[height=2.5in]{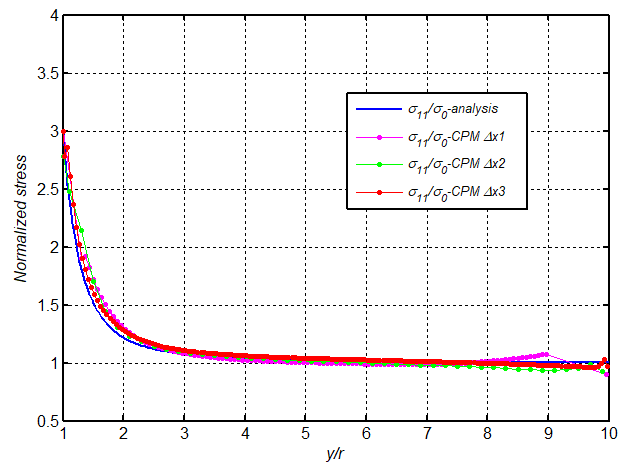}
	\caption{Comparison of the CPDM solution and the analytical Kirsch solution.}
\label{fig:com-analysis}
\end{figure}

The geometry and boundary setting are shown in Fig. \ref{fig:phsetting}.
The side length of the square plate is 100 mm,
and a circular hole with a radius of 5 mm is opened in the center of the square plate.
In the numerical simulations,the spacing between particles
is $5mm$.
In addition, in order to verify the convergence of CPDM,
we also use this example to compare CPDM results
with analysis values.
The analytical Kirsch's solution of stress components around a circular
hole in an elastic infinite plate under tension are give as follows,
\begin{eqnarray}
	\sigma_{rr}&=&
\frac{\sigma_0}{2}(1-\frac{a^2}{r^2})
+\frac{\sigma_0}{2}(1-\frac{a^2}{r^2})(1-3\frac{a^2}{r^2})\cos2\theta
\\
	\sigma_{\theta \theta}&=&\frac{\sigma_0}{2}
(1+\frac{a^2}{r^2})-\frac{\sigma_0}{2}(1+3\frac{a^4}{r^4})\cos2\theta
\\
	\sigma_{r \theta} &=&
\sigma_{\theta r }=-\frac{\sigma_0}{2}(1-\frac{a^2}{r^2})(1+3\frac{a^2}{r^2})sin2\theta
\end{eqnarray}
where $\theta$ and $r$ are
the polar coordinates measured form the center of the circular hole.
$r$  is the radial distance of the point of interests
to the center of the hole, and $\theta$ is the angle
between the x-axis and the radial vector ${\bf r}$ as shown in Fig. 17.
The comparison between the CPDM solution and the analytical Kirsch's
solution given shown in Fig.\ref{fig:com-analysis}.

\begin{figure}[H]	
	\centering
	\begin{subfigure}[t]{0.3\linewidth}
		\centering
		\includegraphics[height=1.4in]{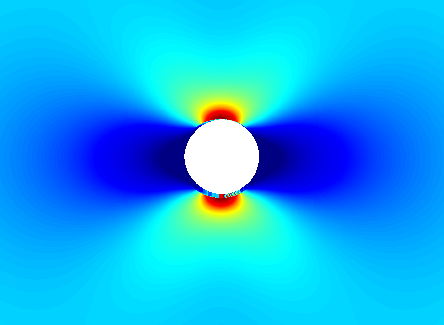}
		\caption{S11}
	\end{subfigure}
	\begin{subfigure}[t]{0.3\linewidth}
		\centering
		\includegraphics[height=1.4in]{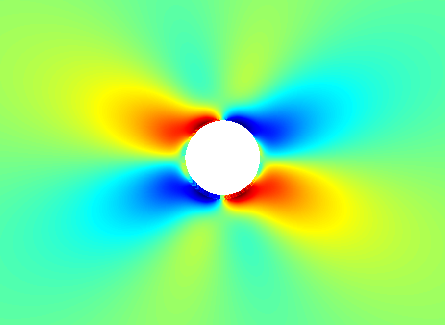}
		\caption{S12}
	\end{subfigure}
	\begin{subfigure}[t]{0.3\linewidth}
		\centering
		\includegraphics[height=1.4in]{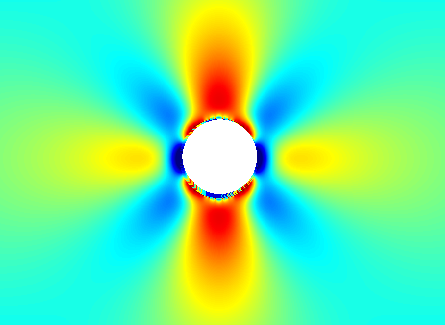}
		\caption{S22}
	\end{subfigure}
	\caption{stress distribution near the hole}
\label{fig:phstress22}
\end{figure}

This example not only validate CPDM method, but also validate and verify
the cohesive stress formulation derived in this paper.
The variation of stress distribution is shown in Fig. \ref{fig:phstress22}.


\subsection{Double-edge notched specimen test}
In this example, we conducted a numerical simulation of
a double-edge notched specimen under tensile loading, which was based
on the experiment reported in \cite{2005Crack}.
In the original experiment,
the notch was set to be asymmetric, and hence it is a mixed-mode fracture.
This example provides us an opportunity to test how CPDM to handle mixed mode
fracture.

\begin{figure}[H]
	\centering
	\includegraphics[height=2.5in]{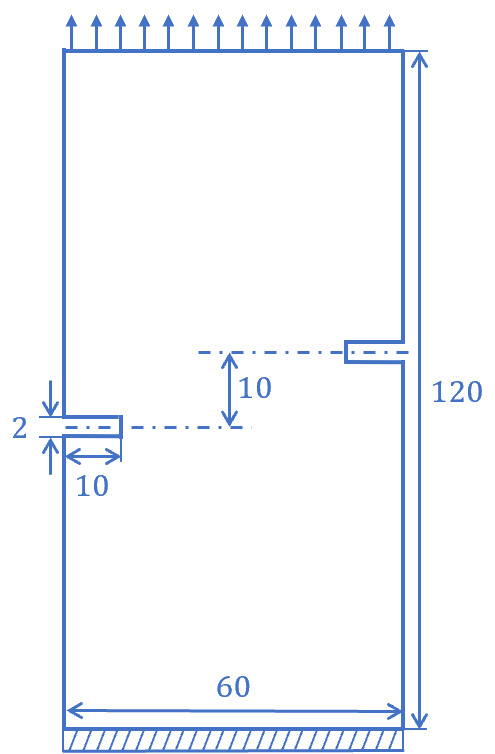}
	\caption{Sketch map of double edge notched tensile test}\label{fig:doubledge}
\end{figure}

The problem setting is displayed in the Fig. \ref{fig:doubledge}.
The numerical specimen is $120 mm$ long and $60 mm$ wide.
Two notches are set respectively at two lateral sides with $5 mm$ from the middle
line from above and below.
Each notch is $10mm$ long and $2 mm$ wide.
The material parameters of the specimen are: $E = 40 GP_a$ , $\mu = 0.2$ , and $G_{f} = 0.025 N/mm$.
The particle spacing in the numerical simulation is chosen as $5 mm$, and the horizon size
is taken as $3.015$ times the particle spacing.
\begin{figure}[H]	
	\centering
	\begin{subfigure}[t]{0.23\linewidth}
		\centering
		\includegraphics[height=1.8in,width=1.2in]{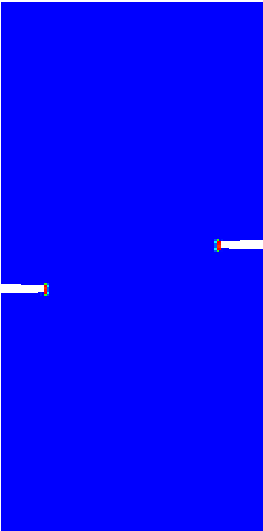}
		\caption{t=0.66ms}\label{fig:dmgmixmode1}		
	\end{subfigure}
	\begin{subfigure}[t]{0.23\linewidth}
		\centering
		\includegraphics[height=1.8in,width=1.2in]{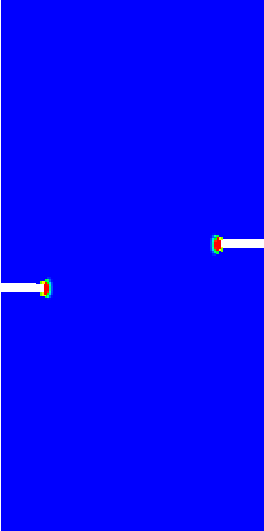}
		\caption{t=0.715ms}\label{fig:dmgmixmode2}
	\end{subfigure}
	\begin{subfigure}[t]{0.23\linewidth}
		\centering
		\includegraphics[height=1.8in,width=1.2in]{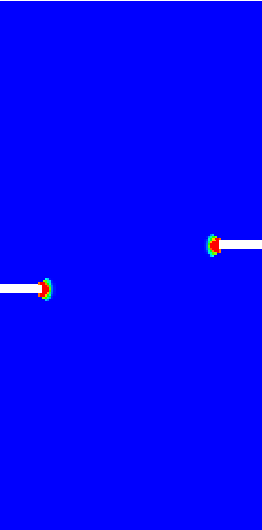}
		\caption{t=0.77ms}\label{fig:dmgmixmode3}
	\end{subfigure}
	\begin{subfigure}[t]{0.05\linewidth}
		\centering
		\includegraphics[height=0.8in]{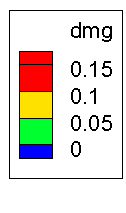}
	\end{subfigure}
	\hfill
	\begin{subfigure}[t]{0.23\linewidth}
		\centering
		\includegraphics[height=1.8in,width=1.2in]{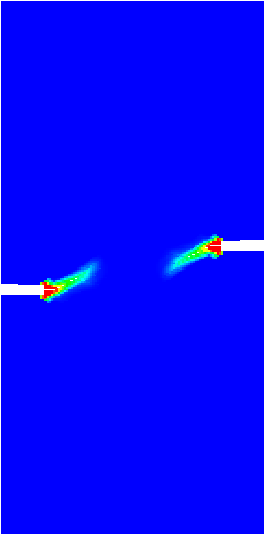}
		\caption{t=0.825ms}\label{fig:dmgmixmode4}
	\end{subfigure}
	\begin{subfigure}[t]{0.23\linewidth}
		\centering
		\includegraphics[height=1.8in,width=1.2in]{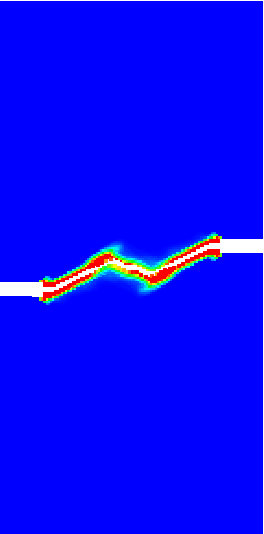}
		\caption{t=0.88ms}\label{fig:dmgmixmode5}
	\end{subfigure}
	\begin{subfigure}[t]{0.23\linewidth}
		\centering
		\includegraphics[height=1.8in,width=1.2in]{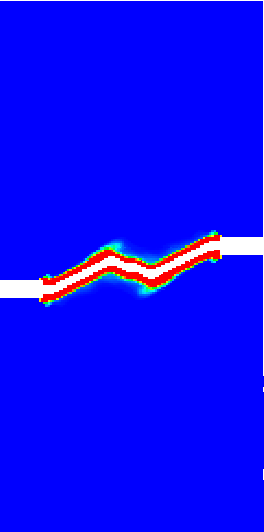}
		\caption{t=0.935ms}
     \label{fig:dmgmixmode6}
	\end{subfigure}
	\begin{subfigure}[t]{0.05\linewidth}
		\centering
		\includegraphics[height=0.8in]{dmgmixmode.png}
	\end{subfigure}
	\caption{Damage distribution at different time instances}
\label{fig:mixmodedamage}
\end{figure}
Figure \ref{fig:mixmodedamage} shows the sequence of crack growth with the damage
contour at different time instances, while
Fig. \ref{fig:mixmodestress} displays the sequence of crack growth with the stress $S_{22}$
contour.
\begin{figure}[H]	
	\centering
	\begin{subfigure}[t]{0.25\linewidth}
		\centering
		\includegraphics[height=1.8in,width=1.2in]{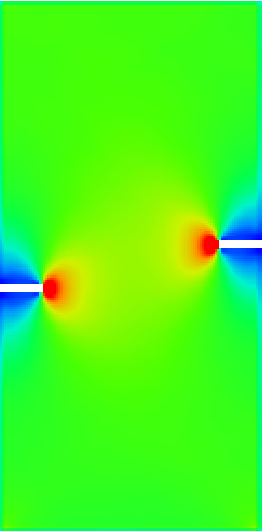}
		\caption{t=0.66ms}\label{fig:mixmode2}		
	\end{subfigure}
	\begin{subfigure}[t]{0.25\linewidth}
		\centering
		\includegraphics[height=1.8in,width=1.2in]{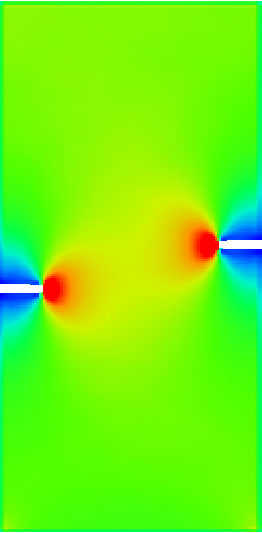}
		\caption{t=0.715ms}\label{fig:mixmode3}
	\end{subfigure}
	\begin{subfigure}[t]{0.25\linewidth}
		\centering
		\includegraphics[height=1.8in,width=1.2in]{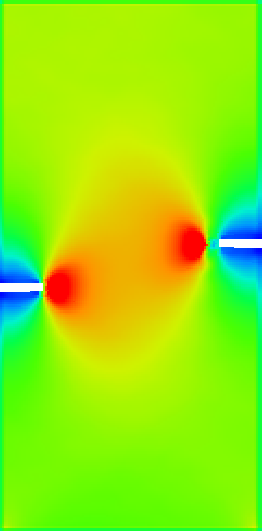}
		\caption{t=0.77ms}\label{fig:mixmode4}
	\end{subfigure}
	\begin{subfigure}[t]{0.05\linewidth}
	\centering
	\includegraphics[height=1.35in,width=0.9in]{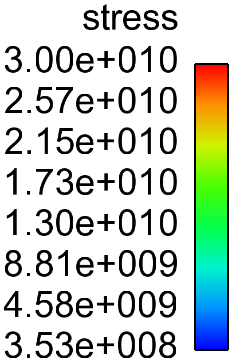}
    \end{subfigure}
	\hfill
	\begin{subfigure}[t]{0.25\linewidth}
		\centering
		\includegraphics[height=1.8in,width=1.2in]{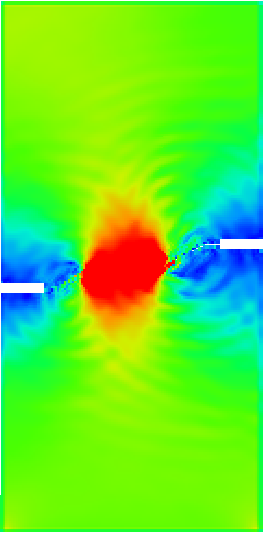}
		\caption{t=0.825ms}\label{fig:2stress128}
	\end{subfigure}
	\begin{subfigure}[t]{0.25\linewidth}
		\centering
		\includegraphics[height=1.8in,width=1.2in]{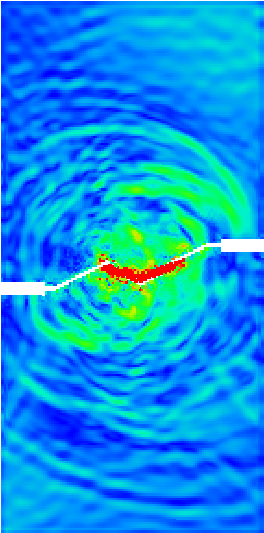}
		\caption{t=0.88ms}\label{fig:mixmode6}
	\end{subfigure}
    \begin{subfigure}[t]{0.25\linewidth}
    	\centering
    	\includegraphics[height=1.8in,width=1.2in]{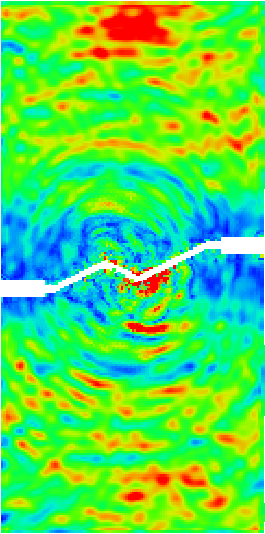}
    	\caption{t=0.935ms}\label{fig:mixmode7}
    \end{subfigure}
	\begin{subfigure}[t]{0.05\linewidth}
	\centering
	\includegraphics[height=1.35in,width=0.9in]{mixmode_color.png}
    \end{subfigure}
	\caption{Stress distribution at several typical moments}
\label{fig:mixmodestress}
\end{figure}

\subsection{Three point bending test}
\begin{figure}[H]
	\begin{center}
		\includegraphics[height=2.0in]{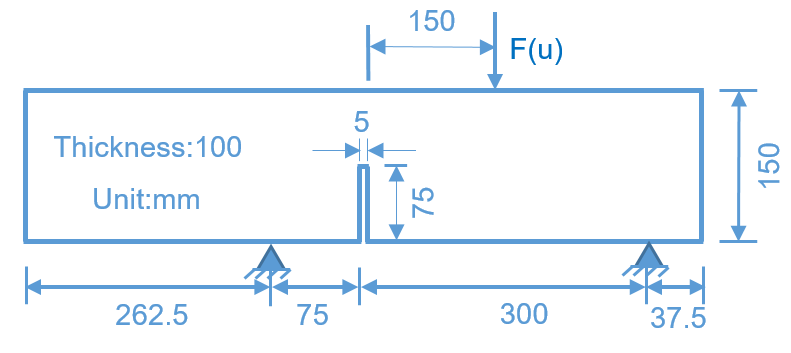}
	\end{center}
	\caption{Sketch map of three point bending test}
	\label{fig:TPBsetting}
\end{figure}

\begin{figure}[H]
	\begin{center}
		\centering
		\includegraphics[height=1.2in]{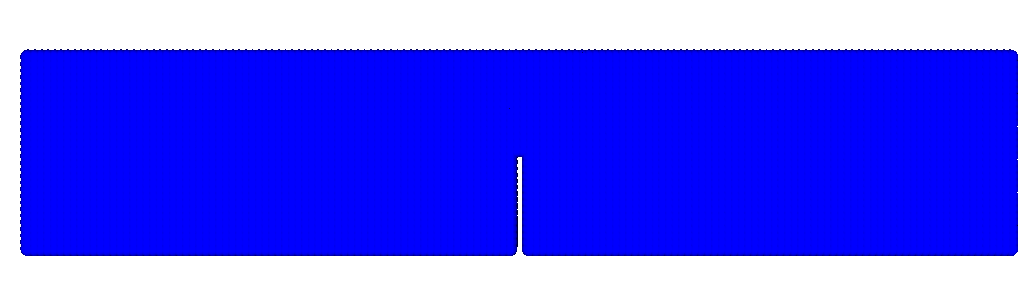}
		\label{fig:Threepointbending3}
	\end{center}
	\begin{center}
		\centering
		\includegraphics[height=1.8in]{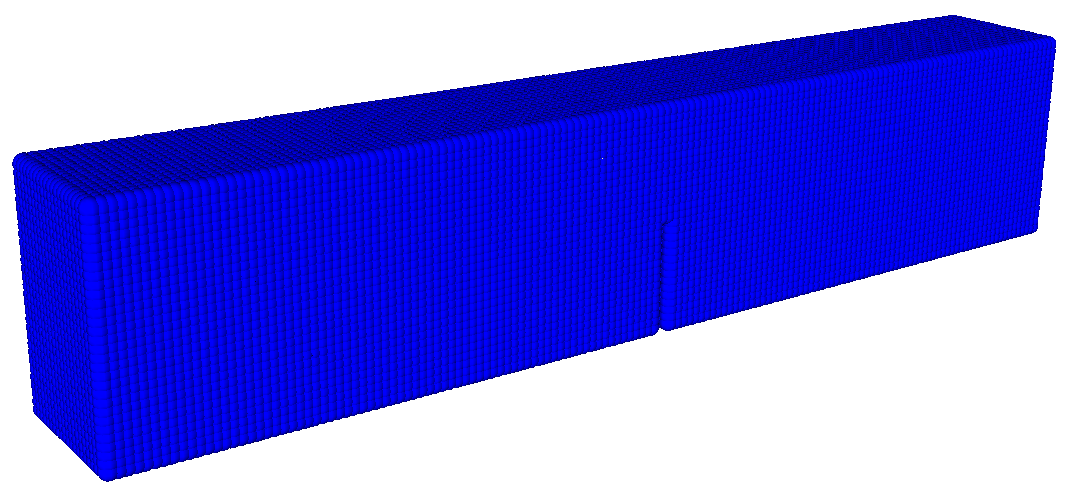}
		\label{fig:Threepointbending2}-
	\end{center}
	\caption{Numerical model of three point bending test}
	\label{fig:3D-example-setup}
\end{figure}

The third example is a three-dimensional (3D) simulation
of cohesive peridynamics, in which we conducted a numerical test
of the three-point bending of a beam \cite{JC1998Mixed}.
This example is widely used as a benchmark problem of 3D mixed mode fracture.
The simulation domain and boundary condition are shown in Fig.\ref{fig:TPBsetting}.
The material properties are set as follows:
the Young' modules $E = 20 GP_a$; Poisson's ratio $\nu$ = $0.2$;
and the fracture energy $G_f = 0.015 N/mm$.
A concentrated downward load is acted at
the upper midpoint of the beam,
and the left and right ends of the lower part are fixed by the fixative structure.
In the numerical simulations,
the spacing between particles is chosen $5mm$.
The 3D numerical specimen is shown in Fig.\ref{fig:3D-example-setup}.

\begin{figure}[H]	
\begin{minipage}{1\linewidth}
		\begin{subfigure}[t]{0.76\linewidth}
		\centering
		\includegraphics[height=1.1in]{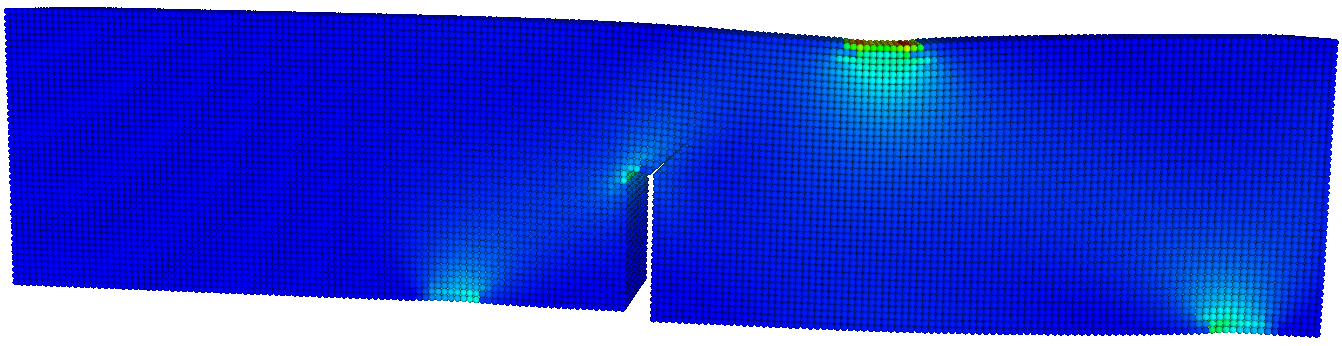}
	\end{subfigure}
		\begin{subfigure}[t]{0.15\linewidth}
	\centering
	\includegraphics[height=1.0in]{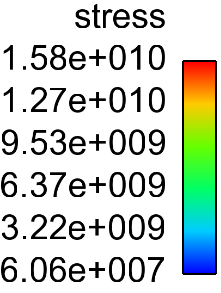}
\end{subfigure}
\begin{center}
(a)
\end{center}
\end{minipage}
\begin{minipage}{1\linewidth}
		\begin{subfigure}[t]{0.76\linewidth}
	\centering
	\includegraphics[height=1.1in]{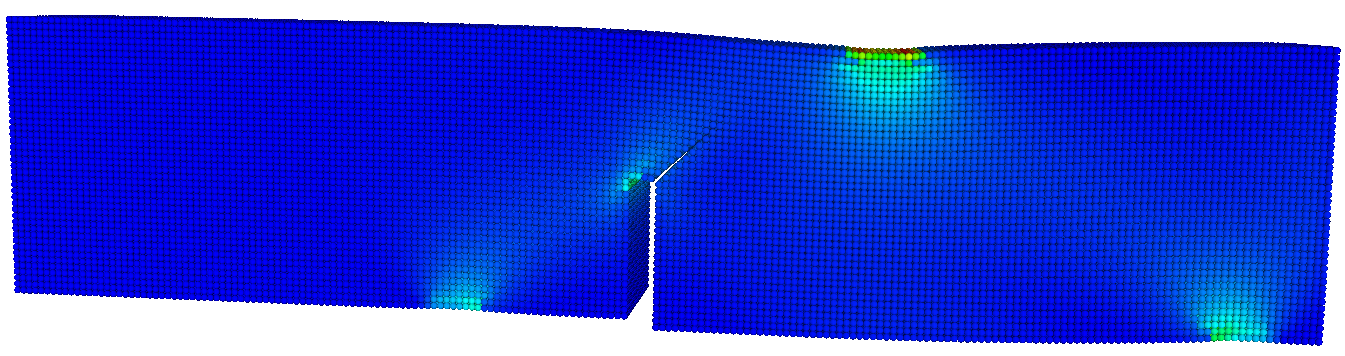}
\end{subfigure}
\begin{subfigure}[t]{0.15\linewidth}
	\centering
	\includegraphics[height=1.0in]{TPB_stress_color.png}
\end{subfigure}
\begin{center}
(b)
\end{center}
\end{minipage}
\begin{minipage}{1\linewidth}
	\begin{subfigure}[t]{0.76\linewidth}
	\centering
	\includegraphics[height=1.15in]{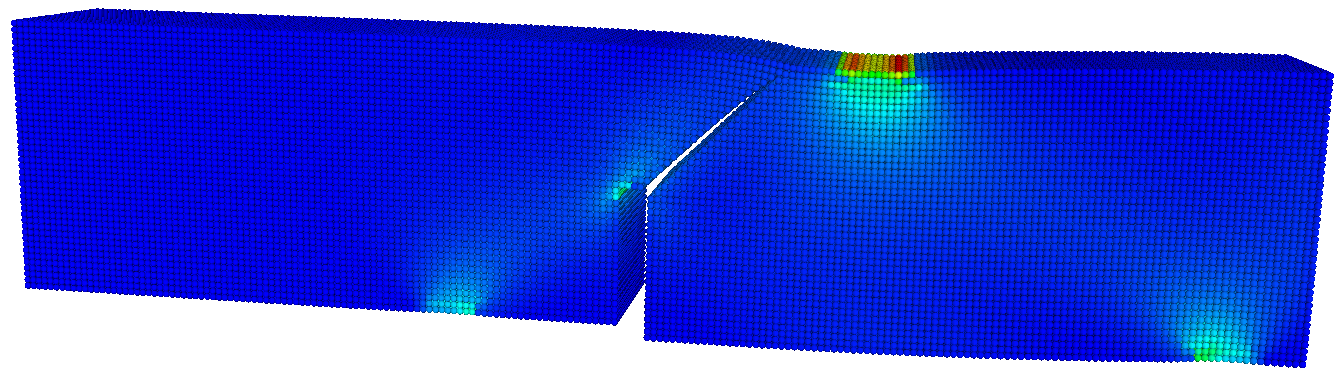}
\end{subfigure}
\begin{subfigure}[t]{0.15\linewidth}
	\centering
	\includegraphics[height=1.0in]{TPB_stress_color.png}
\end{subfigure}
\begin{center}
(c)
\end{center}
\end{minipage}
	\caption{Von Mises stress distribution at (a) $t=5ms$, (b) $t=6ms$, and (c) $t=7ms$,}
	\label{fig:stressDistrib}
\end{figure}

\begin{figure}[H]
\begin{minipage}{1\linewidth}
	\begin{subfigure}[t]{0.75\linewidth}
	\centering
	\includegraphics[height=1.2in]{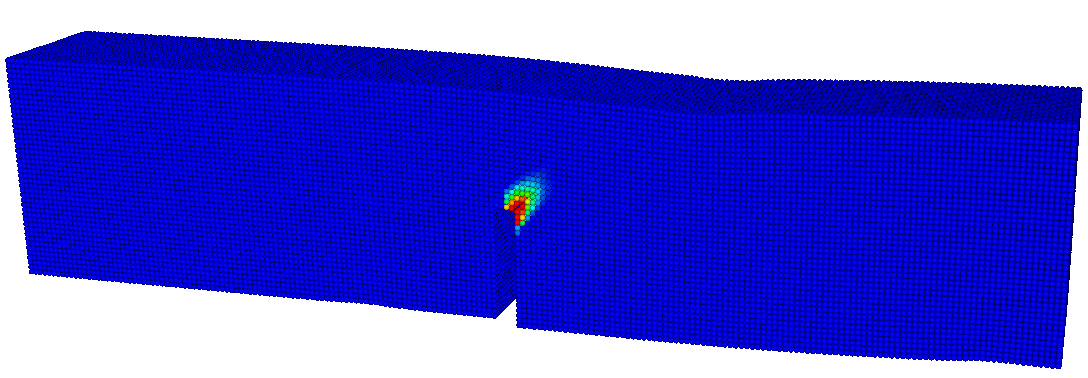}
\end{subfigure}
\begin{subfigure}[t]{0.15\linewidth}
	\centering
	\includegraphics[height=1.0in]{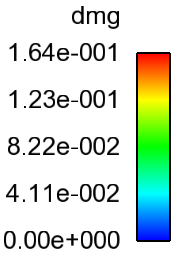}
\end{subfigure}
\begin{center}
(a)
\end{center}
\end{minipage}
\begin{minipage}{1\linewidth}
	\begin{subfigure}[t]{0.75\linewidth}
	\centering
	\includegraphics[height=1.13in]{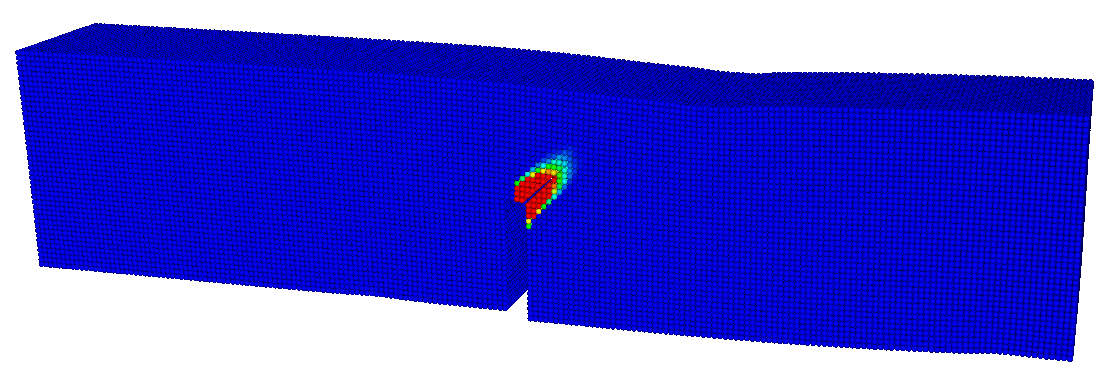}
\end{subfigure}
\begin{subfigure}[t]{0.15\linewidth}
	\centering
	\includegraphics[height=1.0in]{TPB_dmg_color.png}
\end{subfigure}
\begin{center}
(b)
\end{center}
\end{minipage}
\begin{minipage}{1\linewidth}
	\begin{subfigure}[t]{0.75\linewidth}
	\centering
	\includegraphics[height=1.13in]{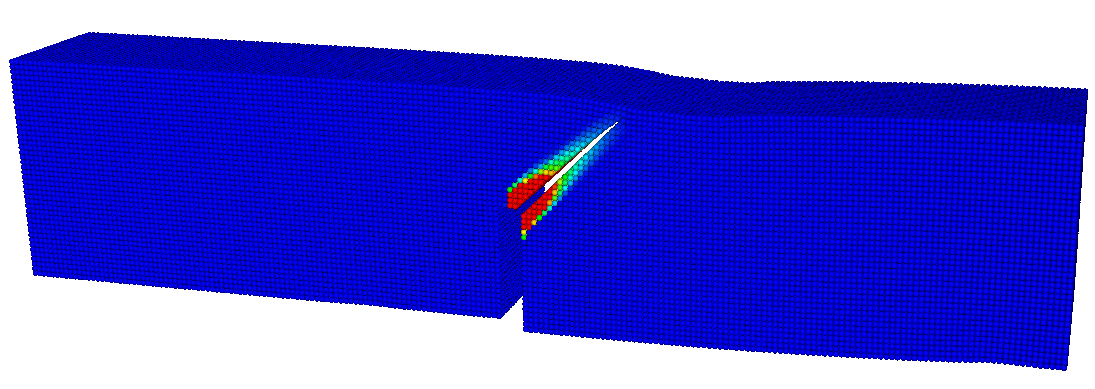}
\end{subfigure}
\begin{subfigure}[t]{0.15\linewidth}
	\centering
	\includegraphics[height=1.0in]{TPB_dmg_color.png}
\end{subfigure}
\begin{center}
(c)
\end{center}
\end{minipage}
	\caption{damage distribution at (a) $t=5 ms$, (b) $t=6 ms$, and (c) $t=7 ms$.}
	\label{fig:TPBdmg}
\end{figure}

We plot the fracture process of the simply supported beam in Fig.\ref{fig:stressDistrib}.
Figure \ref{fig:stressDistrib} (a), (b), and (c)
show the stress distribution of the beam at different time instances during the loading,
and Fig.\ref{fig:TPBdmg} (a), (b), and (c) shows the damage evolution with respect to the time.

\begin{figure}[H]
	\begin{center}
		\includegraphics[height=3.0in]{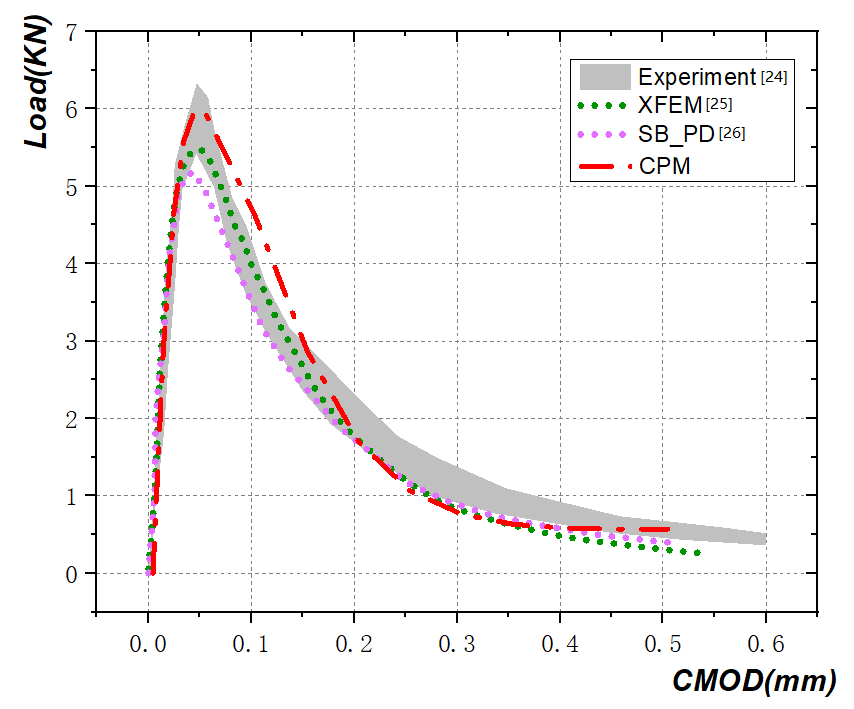}
	\end{center}
	\caption{Crack mouth opening displacement-force relations for the three point bending tests:
	Experiment data \cite{2010Application};
	the eXFEM simulation data  \cite{2007Modelling};
	a previous peridynamics (state based peridynamics) simulation data \cite{Yang2019}, and
    the data from the present work (CPDM) .
}
 \label{fig:dataref}
\end{figure}

Fig.\ref{fig:dataref} shows the change of force with respect to crack mouth opening displacement (CMOD).
In Fig.\ref{fig:dataref}, the dotted line represents the numerical calculation result obtained by using
the cohesive peridynamics and the gray area shows the experiment data,
the other lines present the results obtained by using the extended finite element (eXFEM)
and the state-based peridynamics that is mislabeled as the cohesive zone peridynamics method.

\section{Discussions and conclusions}

It has been almost four decades since Xu and Needleman's
pioneer work on cohesive zone model (CZM) \cite{Xu1994},
and finite element CZM method established itself as a primary numerical method
to model inelastic fracture at small scale yielding.

However, CZM has some technical issues such as mesh dependence, adaptive mesh refinement, and
modeling on mixed mode fracture, among others.
Among these shortcomings, one of the major criticisms on CZM is its replacement of a homogeneous
material by a cohesive interface ``composite'' in its core mechanics model.
On the other hand,
the cohesive peridynamics model (CPDM) proposed in this work is a nonlocal cohesive continuum mechanics,
and it has demonstrated its potential to address all the issues mentioned above, which
exist in finite element cohesive zone model.
Moreover, unlike the prototype micro-brittle (PMB) model in the bond-based peridynamics,
the bond-based cohesive peridynamics model (CPDM) has variable Poisson's ratio, does not need
the critical stretch parameter $s_0$, and provides intrinsic stress measure that is consistent
with strain measure in the bulk homogeneous continua.
Thus, CPDM is a bona fide nonlocal cohesive continuum mechanics modeling.

Furthermore, in this work, we have shown in the first time
that the elastic tensor derived from the mesoscale pairwise
Xu-Needleman potential is not limited by the Cauchy relation.
This stunning discovery will greatly broaden applications of the bond-based
peridynamics.

It should be mentioned that in the literature some authors labeled their state-based peridynamics
or coupling method between finite element method and the state-based peridynamics
as the cohesive peridynamics, e.g. \cite{Tong2020,Yang2019,Yang2020}.
Even though there are some cohesive zone features in these state-based peridynamics,
they are not really cohesive peridynamics model, because in those work
the micrscale cohesive bond force models do not govern the macroscale constitutive relations.
Whereas, in the proposed CPDM model the mesoscale or microscale cohesive bond potential
determines the material constitutive relations at macroscale.
Therefore, CPDM is a consistent cohesive zone model.

\bigskip
\bigskip

\noindent
\centerline{\bf Acknowledgments}
\\

The authors would like to thank Dr.Xuan Hu of the University of California,
Berkeley for some helpful discussions.
J.H and A.Z are supported by the National Natural Science Foundation of China
(Grant number 52088102 and 51925904), which are gratefully acknowledged.

\section*{References}
\bibliographystyle{unsrt}
\bibliography{ref}

\end{sloppypar}
\end{document}